\documentclass{aa}

\usepackage{graphicx}
\usepackage{txfonts}
\usepackage{hyperref}
\begin{document}

   \title{The 2024 KIDA network for interstellar chemistry}


   \author{V. Wakelam
          \inst{1}
          \and
          P. Gratier\inst{1}
          \and
          J.-C. Loison\inst{2}
           \and
           K. M. Hickson\inst{2}
        \and
        J. Penguen\inst{3}
        \and
          A. Mechineau\inst{1}
          }

   \institute{Laboratoire d’astrophysique de Bordeaux, Univ.Bordeaux, CNRS, B18N, allée Geoffroy Saint-Hilaire, 33615 Pessac, France\\
              \email{valentine.wakelam@u-bordeaux.fr}
         \and
         Institut des Sciences Moléculaires (ISM), CNRS, Univ.Bordeaux, 351 cours de la Libération, 33400 Talence, France
         \and
         Observatoire Aquitain des Sciences de l'Univers, Univ.Bordeaux, CNRS, B18N, allée Geoffroy Saint-Hilaire, 33615 Pessac, France\\
             }

   \date{Received xxx; accepted xxx}

 
  \abstract
   {The study of the chemical composition of the interstellar medium (ISM) requires a strong synergy between laboratory astrophysics, modeling, and observations. In particular, astrochemical models have been developed for decades now and include an increasing number of processes studied in the laboratory or theoretically. These models follow the chemistry both in the gas phase and at the surface of interstellar grains. Since 2012, we have provided complete gas-phase chemical networks for astrochemical codes that can be used to model various environments of the ISM. }
   {Our aim is to introduce the new up-to-date astrochemical network kida.uva.2024 together with the ice chemical network and the fortran code to compute time dependent compositions of the gas, the ice surface, and the ice mantles under physical conditions relevant for the ISM. }
   {The gas-phase chemical reactions, as well as associated rate coefficients, included in kida.uva.2024 were carefully selected from the KIDA online database and represent the most recent values. The model predictions for cold core conditions and for when considering only gas-phase processes were computed as a function of time and compared to the predictions obtained with the previous version, kida.uva.2014. In addition, key chemical reactions were identified. The model predictions, including both gas and surface processes, were compared to the molecular abundances as observed in the cold core TMC1-CP. }
   {Many gas-phase reactions were revised or added to produce kida.uva.2024. The new model predictions are different by several orders of magnitude for some species. The agreement of this new model with observations in TMC-1 (CP) is, however, similar to the one obtained with the previous network.}

\titlerunning{KIDA 2024}
   \maketitle
%

\section{Introduction}

New observational capabilities have allowed for fast mapping of the interstellar medium (ISM) while observing over a larger spectral range. As such, many new regions are being chemically characterized, and new molecules are also being discovered. In order to interpret and even prepare these observations, astrochemical models are being developed. Pure gas-phase models are no longer used to study the cold or warm dense regions of the ISM because grain chemistry plays an essential role in the determination of the gas-phase composition of these regions. Although many uncertainties remain in the surface processes and the gas-surface interactions, it is not possible to ignore these gas-grain interactions. In recent years, experimental and theoretical studies have been undertaken to determine with a better accuracy some of the key parameters, such as binding energies, for surface chemistry \citep[see for instance the review by ][]{2022ESC.....6..597M}.

Astrochemical models are a combination of a numerical code (hereafter called the astrochemical code), chemical networks (a list of reactions but also other types of chemical parameters, such as binding energies), and a physical model (the time dependent or fixed physical parameters) adequate for an astrophysical object. In many cases, these three parts are not independent, meaning that processes are selected to fit one specific type of object, so the code and chemical networks cannot be used for other objects. 
Many gas-grain astrochemical codes and chemical networks exist in the literature \citep[see for instance the review of chemical models for protoplanetary disks by][]{2013ChRv..113.9016H}. Most of them are not public, and the results published in the literature are difficult to compare. Among the few public codes, we can cite UCLCHEM\footnote{https://uclchem.github.io/} \citep{2017AJ....154...38H}, CHEMPL\footnote{https://github.com/fjdu/chempl} \citep{2021RAA....21...77D}, and Nautilus\footnote{https://astrochem-tools.org/codes/} \citep{2016MNRAS.459.3756R}. Only a few chemical networks are public and documented at a minimum level. The UMIST Database for Astrochemistry (UDfA) has been available for a long time \citep{1991A&AS...87..585M,1997A&AS..121..139M,2000A&AS..146..157L,2007A&A...466.1197W,2013A&A...550A..36M, 2024A&A...682A.109M}. The chemical networks developed by the team of Professor Eric Herbst \citep[known as the OSU network,][and references therein]{2010ApJ...721.1570H} had been online for many years before becoming the KIDA database \citep{2012ApJS..199...21W,2015ApJS..217...20W}. Recently, a new gas-phase chemical network has been made available online: the GRETOBAPE gas-phase reaction network \citep{2023ApJS..266...38T}.  Even though ice chemistry is essential, the chemical model predictions will always depend on the gas-phase network. 

Among the thousands of gas-phase chemical reactions used in these models, most of them have not been studied experimentally or theoretically, especially under conditions relevant to the ISM \citep{2010SSRv..156...13W}. In many cases, the rate coefficients and/or branching ratios of the products have been determined using common sense or chemical intuition, sometimes guided by information about chemically related reactions. In 2009, we set up the KInetic Database for Astrochemistry,\footnote{https://kida.astrochem-tools.org/} which is an online database collecting chemical reactions (with rate coefficients and bibliographic references) published in the literature and relevant for ISM chemistry \citep{2012ApJS..199...21W}. In 2016, we extended the database to store chemical reactions at the surface of the grains, in addition to species binding and diffusion energies. From this online database, we extracted a sub-list of reactions in the form of a chemical network (kida.uva) that can be directly used with the astrochemical code we provide (namely, Nautilus) or adapted for any other astrochemical code. Here "uva" stands for University of Virginia, as the initial data fed to the database originate from the gas-phase chemical networks developed by Professor Eric Herbst (OSU database). In 2012, we delivered the first version of this network \citep[kida.uva.2011,][]{2012ApJS..199...21W}, and the first upgraded version was released in 2015 \citep[kida.uva.2014,][]{2015ApJS..217...20W}. Since then, a large number of reactions have been updated in the online KIDA database. In this paper, we present the new 2024 version of the kida.uva chemical network, which includes all the updates up to January 2024. In addition, we present a new version of the Nautilus astrochemical model and a surface chemical network compatible with this gas-phase network. As such, we provide the community with a full up-to-date open access tool to simulate both the gas and ice composition under interstellar conditions.

\section{The new kida.uva.2024 gas-phase network}

\subsection{Method to produce the network}

\begin{table*}
    \centering
     \caption{Identification number for reaction types listed in the chemical file.}
    \begin{tabular}{ll}
    \hline
    \hline
      ID   & Type of reaction \\
    \hline
       0  & Ion recombination with grains \\
       1  & Photodissociation/ionzation with cosmic-rays (CR) \\
       2 & Gas phase photodissociations/ionzations by secondary UV photons generated by CR \\
       3  & Gas phase photodissociations/ionzations by UV \\
       4 &      Neutral-neutral, ion-neutral, anion-cation reactions and associative ionization \\
       5 &      Exchange reactions \\
       6 & Association reactions between two species (neutral or ionized) stabilized by the emission of a photon \\
       7 & Association of a neutral species and an anion, resulting in the ejection of the extra electron \\
       8 & Recombination of a positive ion with an electron resulting in the dissociation of the molecule or the \\
        & emission of a photon or the attachment of an electron \\
       10-11 & Forcing H$_2$ formation \\
       14 & Grain surface reactions \\
       15 & Thermal evaporation \\
       16 & Evaporation induced by cosmic-ray stochastic heating \\
       17-18 & Photodissociations by cosmic-ray induced UV photons on grain surfaces  \\
       19-20 & Photodissociations by UV photons on grain surfaces \\
       40 & Swapping reactions of surface species toward the mantle  \\
       41 & Swapping reactions of mantle species toward the surface \\
       66 & Photodesorption by external UV \\
       67 & Photodesorption by UV induced by cosmic-rays \\
       77 & Grain sputtering  \\
       99 & Adsorption on grains \\
    \hline
    \end{tabular}
    \label{tab:id_types}
\end{table*}

The KIDA online database contains kinetic data for gas-phase reactions and diffusion reactions at the surface of the grains. 
The database stores a number of reactions (with rate coefficients and species parameters (such as desorption and diffusion energies). A fraction of these data are associated with a bibliographic reference. The rest of the data without a proper reference have been added through chemical intuition. Educated guesses have been made regarding the products and rate coefficients of these processes. For some data, several values are included in KIDA due to different origins. Even data with references can be uncertain for various reasons, for example, several groups have derived inconsistent results or the data have been computed or measured in conditions different from the ones that are required. Keeping KIDA updated is a difficult task, and it is done in parallel to our research. As such, creating this kida.uva.2024 network was the occasion to update the KIDA database with reactions added in our private networks. To create the sub network kida.uva.2024, we first extracted all the gas-phase reactions stored in KIDA and applied a multi-step process, which we explain below.\\

\subsubsection{Reactions with multiple rate coefficients in KIDA}

For the reactions with several rate coefficients in KIDA, the first task was to select a unique value for a range of temperature. The selection was usually done by choosing the most recent value. 

\subsubsection{Temperature ranges of validity}
By default, especially when there was no value mentioned in the original paper, the temperature range of validity was set as 10 to 800~K. For some rate coefficients, there is a specific range (due to the experimental or theoretical conditions). For other reactions, there are several rate coefficients given for complementary ranges of temperatures. This range of temperature is important for reactions with temperature dependent rates because one needs to decide what happens outside of the range. For reactions with a minimum temperature range significantly above 10 K, we checked their impact on the chemical predictions, and sometimes we changed it. For instance, the reaction N + H$_2$ $\rightarrow$ H + NH reaction is defined over a range of 1950 - 2850 K in KIDA based on the experimental work of \citet{Davidson1990}, with a rate coefficient of $\rm k(T) = 2.66\times 10^{-10} \exp{(-12600/T)}$. For a simulation at 10 K, if this rate was computed at 1950 K, the rate coefficient would be $4.16\times 10^{-13}$~cm$^{-3}$~s$^{-1}$. Considering the high abundances of the reactants, this would produce large amounts of NH that would recombine with C$^+$ and influence the abundances of many species. Since the reaction has a strong activation barrier, it is very unlikely that the rate coefficient at 10 K is so large. As a consequence, we modified the range of temperature for this reaction to 10 - 2850 K in the kida.uva.2024 network. This ensures a negligible rate at low temperatures.
\subsubsection{Completeness of the network}
We checked that all the species had both production and destruction reactions since this is not a prerequisite for inclusion of data in the KIDA online database. For species for which we had either no production or no destruction pathway, we did not include them in kida.uva.2024. These species are given in Appendix~\ref{removed-from-KIDA} for information. If one is interested in these species, they have to download the existing reactions from KIDA, complete their chemistry, and add them in the kida.uva.2024 network. 

We also compared the species and reactions in kida.uva.2024 with those in the network we maintain and develop in our group. We have found a number of species and reactions either linked to the surface processes (such as C$_n$H$_m$ species) or from specific works (such as S$_n$ species), which were absent from kida.uva.2024. These species and reactions have been added to the KIDA online database (with reference to this paper or the original paper when relevant) and to the kida.uva.2024 network (see Appendix~\ref{additions-from-KIDA}). 

We checked that all neutral species have a destruction reaction with cosmic-ray induced photons and direct UV photons and that all cations have a recombination reaction with electrons. When missing, we added such reactions. We note that the assumed rates and products are highly uncertain. In some cases, we have used existing values from similar systems or a default reasonable value. The philosophy behind this is that it is less wrong to add such reactions with a very uncertain rate coefficient (and products) rather than not include them at all. The list of such reactions is given in Table~\ref{tab:added-reactions}, and these have now been added to the KIDA online database with an "Not rated value" flag and a reference to this paper. 
\subsubsection{Check of endothermic reactions}
Both the kida.uva.2024 and the online KIDA database have been corrected for the endothermic reactions identified by \citet{2023ApJS..266...38T} in the KIDA online database. We note that some of these reactions were in kida.uva.2014. We identified only one problematic reaction that really impacted the model: CH$_4$ + S$^+$ $\rightarrow$ H +  H$_3$CS$^+$ added from the OSU database and \citet{2017MNRAS.469..435V}. This reaction was producing large quantities of H$_3$CS$^+$ and then H$_2$CS. 

\subsubsection{Comparison with the kida.uva.2014 network}
In order to follow the evolution of the network with respect to the previous version, we have compared the species and reactions listed in the kida.uva.2014 network with the one in our new network. The lists of species removed and added to kida.uva.2024, with respect to kida.uva.2014, are given in Tables~\ref{list-removed-species} and \ref{list-new-species}.

\subsection{Format of the network}

The overall structure of the final files of kida.uva.2024 have not changed compared to the previous version of kida.uva. The gas-phase reaction types are the same as the ones presented in \citet{2015ApJS..217...20W} (ID from 0 to 11 in Table~\ref{tab:id_types}). We briefly recall here the formats of the files. Four files are provided: 
\begin{itemize}
\item gas$\_$species$\_$kida.uva.2024: List of the species with a tabular definition of their elemental composition.
\item gas$\_$reactions$\_$kida.uva.2024: List of chemical reactions with associated parameters.
\item kida.uva.2024.bib: List of bibliographic references used to construct the network in bibtex format.
\item kida.uva.2024$\_$readme.txt: Explanation on the formats of the species and reaction files.
\end{itemize}

The formulas to compute the rate coefficients according to the type of reaction are described in \citet{2012ApJS..199...21W} and in the online database \footnote{https://kida.astrochem-tools.org/help.html}. For reaction types 1, 2, and 3 (from Table~\ref{tab:id_types}), rate coefficients are computed with equations 5, 6, and 7 of \citet{2012ApJS..199...21W}. Reactions types 4 to 8 are computed with equation 1 of \citet{2012ApJS..199...21W}. One exception exists for the reactions between ions and neutral species with a dipole moment for which no measurements exist. The rate coefficients of such processes are computed using the Su–Chesnavich capture approach discussed in  \citet{2009ApJS..185..273W}. Depending on the range of temperature (and the type of formula given in the reaction file), two equations are used to compute the temperature dependent rate coefficients:
\begin{equation}
\rm k(T) = \alpha \beta (0.62+0.4767\gamma(300/T)^{0.5})
\end{equation}
\begin{equation}
\rm k(T) = \alpha \beta (1+0.0967\gamma (300/T)^{0.5} + (\gamma^2/10.526) \times (300/T)).
\end{equation}
Here, $\alpha$, $\beta$, and $\gamma$ are the three parameters given in the gas$\_$reactions$\_$kida.uva.2024 file. {The terms $\alpha$ and $\beta$ are respectively the branching ratio of the reaction and the Langevin rate, while $\gamma$ is the value of the $x$ factor at 300K, as explained in \citet{2012ApJS..199...21W}.\footnote{See also https://kida.astrochem-tools.org/uploads/documents/ionpol$\_$notice.pdf}} Equation 1 is for the type of formula 4 and equation 2 for the type of formula 5\footnote{see also https://kida.astrochem-tools.org/help.html}. 

The rate coefficients for reaction type 0 are described in section 3.2 of \citet{2012ApJS..199...21W}. The temperature dependent rate coefficients for these reactions are computed with equation 1 of \citet{2012ApJS..199...21W} with a null $\gamma$ value. Reactions of the type 10 and 11 are two reactions added to the gas-phase network to mimic the formation of H$_2$ on the grains when the grain surface reactions are not considered. Explanations on these reactions can be found in section 3.2 of \citet{2012ApJS..199...21W}. 

The new network contains 7667 gas-phase chemical reactions to follow 584 species for elements H, He, C, N, O, Si, S, Fe, Na, Mg, Cl, P, and F. The network is available in the ascii format in the KIDA online database.\footnote{https://kida.astrochem-tools.org/networks.html}

\subsection{New gas-phase model predictions}

\begin{table*}[ht]
    \centering
        \caption{Species with significant abundance differences in the model predictions using kida.uva.2014 or kida.uva.2024. }
    \begin{tabular}{lrlrlrlr}
        \hline
        \hline
        Species  & Ratio & Species  & Ratio & Species  & Ratio & Species  & Ratio  \\
        \hline
HCl  &  -0.5 & PN  &  -0.8 & SiC  &  0.5 & SO  &  -0.9\\
CCS  &  -0.5 & CCO  &  0.7 & HNSi  &  0.7 & C$_3$O  &  -0.8\\
C$_3$S  &  -2.3 & H$_2$CO  &  -2.0 & NH$_3$  &  -2.3 & HCCSi  &  -0.9\\
l-C$_3$H  &  -0.7 & C$_6$H$_6$  &  -0.6 & H$_2$CCO  &  -2.7 & c-C$_3$H$_2$  &  -0.5\\
HCNCC  &  -0.5 & HCCNC  &  -0.4 & HNCCC  &  -0.5 & l-C$_3$H$_2$  &  -2.0\\
NC$_4$N  &  -0.5 & C$_4$H$_2$  &  -0.4 & C$_5$H  &  -0.5 & C$_5$N  &  -0.6\\
C$_6$  &  -0.5 & C$_5$O  &  -0.6 & C$_5$H$_2$  &  -0.6 & C$_6$H  &  -0.7 \\
 C$_7$  &  -0.6 & HC$_5$N  &  -0.6 & C$_6$N  &  -0.6 & NC$_6$N  &  -0.5 \\
 C$_6$H$_2$  &  -0.8 & C$_7$H  &  -0.7 & C$_7$N  &  -0.8 & C$_8$  &  -0.6 \\
C$_7$O  &  -0.8 & C$_7$H$_2$  &  -0.8 & C$_8$H  &  -0.7 & C$_9$  &  -0.6 \\
 HC$_7$N  &  -1.0 & C$_8$N  &  -0.8 & C$_8$H$_2$  &  -0.8 & C$_9$H  &  -0.7 \\
 C$_{10}$  &  -0.7 & C$_9$N  &  -0.8 & C$_{10}$H  &  -0.7 & C$_{11}$  &  -0.7 \\
 C$^+$  &  -0.3 & Na$^+$  &  0.3 & S$^+$  &  0.6 & Si$^+$  &  1.5 \\
 CF$^+$  &  -0.4 & SO$^+$  &  -0.5 & CN$^-$  &  -0.6 & CHSi$^+$  &  5.6 \\
 HSiO$^+$  &  -0.7 & H$_2$CCO$_+$  &  -1.1 & H$_2$COH$^+$  &  -1.8 & HC$_3$O$^+$  &  0.8 \\
 HC$_3$S$^+$  &  -0.5 & l-C$_3$H$_2^+$  &  -0.3 & C$_5^-$  &  -0.4 & CH$_3$CO$^+$  &  -0.7 \\
 HC$_3$NH$^+$  &  -0.9 & C$_4$H$_2^+$  &  -0.6 & C$_5$H$^+$  &  -0.4 & H$_2$C$_3$O$^+$  &  3.1 \\ 
 HC$_4$O$^+$  &  -0.6 & C$_5$H$^-$  &  -0.5 & C$_6^-$  &  -0.5 & C$_4$H$_3^+$  &  -0.9  \\
 C$_5$H$_2^+$  &  -0.7 & C$_7^-$  &  -0.6 & C$_7$H$^-$  &  -0.7 & C$_8^-$  &  -0.6 \\
 C$_9^-$  &  -0.6 & & & & \\
        \hline
    \end{tabular}
\\
Species with abundance ratios log(X$_{\rm new}$/X$_{\rm old}$) smaller than -0.3 or larger than 0.3 at 10$^5$ yr and
with an abundance {larger than $10^{-12}$ using the new network} compared with the total H density. The abundances were computed for a standard cold dense core gas-phase only model. (See text for input parameter values.) 
    \label{difference-abundance}
\end{table*}

To estimate the difference in the model predictions using kida.uva.2014 or kida.uva.2024, we first ran a model for the gas-phase reactions only, ignoring any surface chemistry (sticking or surface reactions). We ran the model using both networks for a temperature of 10~K, a density of $2\times 10^4$~cm$^{-3}$~s$^{-1}$, a cosmic-ray ionization rate of $10^{-17}$~s$^{-1}$, and a visual extinction of 15. Then, for the species that are in both networks, we compared the predicted abundances at $10^5$~yr for abundant species (with an abundance larger than $10^{-12}$ using the new network). The species for which the predicted abundances are different by more than a factor of two using the new and old networks --that is, $\rm \log(X_{new}/X_{old})$ smaller than -0.3 or larger than 0.3 -- are listed in Table \ref{difference-abundance}. This list represents about 45\% of the total number of the abundant species at $10^5$ yr. Many of these species are neutral carbon chains (factor of ten difference for HC$_7$N). It is very complicated to identify the reaction or family of reactions responsible for the differences. However, we can take the example of C$_3$S, which is more than two orders of magnitude lower with the new network. This decrease in abundance is due to the addition of the destruction reaction C + C$_3$S $\rightarrow$ C$_3$ + CS from \citet{Vidal2017}. Another example is the large decrease of the NH$_3$ abundance, which is due to the new destruction pathway C + NH$_3$ $\rightarrow$ H$_2$CN + H from \citet{2015ApJ...812..107H}. This molecule, however, is easily produced at the surface of the grains by hydrogenation of atomic nitrogen and desorbed through non-thermal processes in cold environments. As such, the difference in the gas-phase NH$_3$ chemistry is minor when using a full gas-grain code \citep[see also][]{2015ApJ...812..107H}.  


\subsection{Key gas-phase reactions}

While working on astrochemical models, it is useful to identify key reactions. The method usually used first involves looking at the reactions with the largest rates of production or destruction ($\rm R = k_{ij} n_i n_j$ in cm$^{-3}$s$^{-1}$, where $\rm k_{ij}$ is the rate coefficient of the reaction and $\rm n_i$ and  $\rm n_j$ are the densities of the reactants). However, in some cases, the abundance of a species is not sensitive to the exact value of the rate coefficient (usually within a certain range that can be large). We can cite the example of molecular oxygen O$_2$. This molecule was predicted by chemical models to be a reservoir of oxygen in the ISM (after CO), so for a long time it was searched for from space and from the ground, but it was found only in a very few specific places and not in cold cores \citep[see references in][]{2019MNRAS.486.4198W}. Astrochemists identified the production reaction of this species as the O + OH $\rightarrow$ O$_2$ + H reaction, which was assigned a large rate coefficient. New measurements and calculations showed that the rate coefficient was smaller by two orders of magnitude than previously thought \citep{Carty2006,2007JChPh.127b4304X,2008ApJ...681.1318Q}. Despite this, the model predictions were left unchanged because this reaction was also the main destruction reaction for OH, so the overall flux of the reaction was unchanged \citep{2008ApJ...681.1318Q}. For this reason, a reaction that efficiently produces or destroys a species is not necessarily a reaction whose rate coefficient value is going to set its abundance level \citep[see also][]{2010SSRv..156...13W}. To identify key reactions, that is, reactions whose rate coefficients quantitatively influence the species abundances, we used the following sensitivity analysis. For all the gas-phase reactions, we randomly multiplied their rate coefficient by a factor between 0.5 and two independently, creating 4000 different chemical networks. We then ran the pure gas-phase chemical model with these networks, obtaining 4000 different chemical compositions. To identify the key reactions, we computed the Spearman correlation coefficients between each species abundance and each reaction rate coefficient value at each time step. We then identified the reactions that produced a Spearman correlation coefficient larger than 0.3 on species with abundances larger than $10^{-13}$ at two different times: $10^5$ and  $10^6$~yr. We list the reactions that fulfill these criteria for more than one species in Tables~\ref{key-reactions-1e5yr} and \ref{key-reactions-1e6yr}. Common to both times, among the top key reactions, we found the direct ionization of H$_2$ and He by cosmic-ray particles. {We note that the molecular abundances at these times result from the accumulated effect of the reactions up to this time. }
The dependence on these two reactions 
gives an idea of the model's sensitivity to the cosmic-ray ionization rate. Within the lists, one can find all types of reactions (neutral-neutral reactions, ion-neutral reactions, electronic recombination, etc.). Half of the "key" reactions from Table \ref{key-reactions-1e6yr} are also present at an earlier time in Table \ref{key-reactions-1e5yr}. The two tables contain 32 and 22 reactions that can be compared to the 7667 gas-phase reactions of the network. It means that for most of the reactions, a variation of a factor of two in the rate coefficient is not going to significantly affect the abundant species. While compiling these lists, however, we did not take into account the real uncertainty in the rate coefficients. It is possible that some rate coefficients of the listed reactions are better known than a factor of two under ISM cold conditions. Alternatively, it is very likely that some reaction rate coefficients of the network are incorrect by more than a factor of two. In that case, our method would not identify it as a "key" reaction, but changing its value to a better value would impact the model predictions.

\begin{table*}[ht]
    \centering
        \caption{Key reactions {measured} at $10^5$~yr.}
    \begin{tabular}{lc}
        \hline
        \hline
        Reactions  & Impacted species  \\
        \hline
{ H$_2$  +       CR$^*$   $\rightarrow$                  H$_2^+$  +      e$^-$} &  187 \\
{ He     +    CR$^*$       $\rightarrow$              He$^+$   +     e$^-$} & 49 \\
C$_3$       +  CRP$^{**}$ $\rightarrow$                    C     +     C$_2$ & 30 \\
C   +       C$_3$    $\rightarrow$                 C$_4$   +      Photon & CCH, C$_3$, CCO, HCO, OCN, C$_2$H$_2$, c-C$_3$H, C$_3$N, C$_3$O, C$_3$S, l-C$_3$H, c-C$_3$H$_2$,\\
 &  HC$_3$N, HNCCC, l-C$_3$H$_2$, HC$_3$S, HC$_3$O, HC$_4$N, C$_5$O, C$_3$H$^+$, CH$_3^+$, c-C$_3$H$_2^+$, \\
 & HC$_3$O$^+$, SiC$_3$H$^+$, HC$_3$NH$^+$, c-C$_3$H$_3^+$, H$_2$C$_3$O$^+$, l-C$_3$H$_3^+$, H$_2$C$_3$S$^+$, C$_5$H$_3^+$ \\
O     +     CN   $\rightarrow$                  N      +    CO & N, CN, N$_2$, NC$_6$N, NC$_8$N, N$_2$H$^+$ \\
{ H$_2$     +    CH$_3^+$ $\rightarrow$                  CH$_5^+$  +     Photon} & CH$_4$, SiCH$_3$, CH$_2$CCH, C$_4$H$_3$, C$_2$H$_3^+$, CH$_5^+$\\
{ CO    +     H$_3^+$  $\rightarrow$                  H$_2$   +      HCO$^+$} & OH$^+$, H$_2$F$^+$, H$_2$O$^+$, H$_3^+$, H$_3$O$^+$\\
{ H$_2$    +     Si$^+$  $\rightarrow$                  SiH$_2^+$  +    Photon} & SiH, SiS, HNSi, SiH$_2^+$ \\
{ O      +    H$_3^+$ $\rightarrow$                   H$_2$  +       OH$^+$}  & OH$^+$, H$_2$O$^+$, H$_3$O$^+$ \\
c-C$_3$H$_2$ +    Si$^+$ $\rightarrow$                   H    +      SiC$_3$H$^+$ & c-SiC$_2$, l-SiC$_3$, SiC$_3$H$^+$ \\
C   +       NH$_3$    $\rightarrow$                H$_2$CN +      H & NH$_3$, CH$_3$NH$_2$, CH$_3$NH$_3^+$ \\
{ H$_3$O$^+$  +     e$^-$   $\rightarrow$                  H    +      H    +      OH} & H$_2$O, HCOOH, H$_3$O$^+$ \\
{ HF   +      C$^+$  $\rightarrow$                   H  +        CF$^+$}  & HF, CF$^+$ \\
S    +      CH$_3$ $\rightarrow$                   H   +       H$_2$CS & H$_2$CS, HCCS \\
{ CO   +      He$^+$ $\rightarrow$                   He  +       O    +      C$^+$} & He$^+$, NH$_4^+$ \\
N    +      CN     $\rightarrow$                C  +        N$_2$ & N, N$_2$ \\
O    +      CCH  $\rightarrow$                  CH +        CO & CCH, CCO \\
Cl    +     H$_3^+$   $\rightarrow$                 H$_2$  +       HCl+ & Cl, HCl \\
C  +        C$_5$O   $\rightarrow$                 CO  +       C$_5$ & C$_5$O, HC$_5$O$^+$ \\
C    +      H$_2$CO $\rightarrow$                   CO  +       CH$_2$ & HS, H$_2$CO \\
{ Mg  +       HCO$^+$ $\rightarrow$                  HCO +       Mg$^+$} & Mg, Mg$^+$ \\
H$_2$  +       C$_2$H$_2^+$ $\rightarrow$                  C$_2$H$_4^+$ +     Photon & C$_2$H$_2^+$, C$_2$HO$^+$ \\
H$_2$  +       C$_3$H$^+$   $\rightarrow$                l-C$_3$H$_3^+$ +   Photon & C$_3$H$^+$, l-C$_3$H$_3^+$ \\
H$_2$  +       C$_3$H$^+$   $\rightarrow$                c-C$_3$H$_3^+$  +  Photon & C$_3$H$^+$, l-C$_3$H$_3^+$ \\
NH$_3$  +      CH$_3^+$ $\rightarrow$                  CH$_3$NH$_3^+$  +  Photon & CH$_3$NH$_2$, CH$_3$NH$_3^+$ \\
C$_2$H$_4^+$ +     e$^-$      $\rightarrow$               H   +       C$_2$H$_3$ & C$_2$H$_3$, CH$_3$CO \\
HC$_3$NH$^+$ +    e-   $\rightarrow$                  H   +       HNCCC & HNCCC, C$_3$N$^-$ \\
{ CF$^+$  +      e$^-$         $\rightarrow$            C  +        F} & HF, CF$^+$ \\
{ HCS$^+$  +     e$^-$   $\rightarrow$                  S   +       CH} & HCS$^+$, OCS$^+$ \\
SiNC$^+$  +    e$^-$    $\rightarrow$                  Si    +     CN & SiN$^+$, SiNC$^+$ \\
NH$_4^+$    +   e$^-$  $\rightarrow$                   H     +     NH$_3$ & NH$_2$, NH$_4^+$ \\
C$_5$H   +     e$^-$    $\rightarrow$                 C$_5$H$^-$  +     Photon & C$_5$H$_2$, C$_5$H$^-$ \\
        \hline
    \end{tabular}
\\
$^*$ CR means cosmic-ray particle.\\
$^{**}$ CRP means photons induced by cosmic-rays.\\
Boldface reactions are the reactions also listed in Table~\ref{key-reactions-1e6yr}.\\
    \label{key-reactions-1e5yr}
\end{table*}

\begin{table*}[ht]
    \centering
        \caption{Key reactions {measured} at $10^6$~yr.}
    \begin{tabular}{lc}
        \hline
        \hline
        Reactions  & Impacted species  \\
        \hline
{ H$_2$ +        CR$^*$       $\rightarrow$              H$_2^+$ +       e$^-$} &  90 \\
 { CO  +       H$_3^+$     $\rightarrow$               H$_2$ +        HCO$^+$}  & 57 \\
{ He  +       CR$^*$          $\rightarrow$          He$^+$     +   e$^-$} & 54 \\
 { O      +    H$_3^+$     $\rightarrow$        H$_2$  +       OH$^+$} & 34 \\
 H$_2$     +    C$^+$       $\rightarrow$   CH$_2^+$  +     Photon & C$_2$H$_2$, CH$_3$, H$_2$CO, H$_2$CS, H$_2$CCO, CH$_2$NH, CH$_4$, CH$_3$CN, C$_2$H$_4$, CH$_2$CCH$_2$,  \\
 &CH$_3^+$, H$_2$COH$^+$, CH$_3$CO$^+$, CH$_5^+$, CH$_3$NH$_3^+$\\
 O      +    OH      $\rightarrow$                   H     +     O$_2$  & CO, OH, SiN, HNSi, HOOH, NS$^+$ \\
N$_2$     +    He$^+$      $\rightarrow$                  He   +      N    +      N$^+$ & N$^+$, NH$_2^+$, NH$_3^+$, NH$_4^+$ \\
{ H$_2$     +    Si$^+$      $\rightarrow$                  SiH$_2^+$  +    Photon} & SiH, Si$^+$, SiH$^+$, SiH$_2^+$ \\
HCO$^+$ +       e$^-$       $\rightarrow$                  H     +     CO & Mg, Mg$^+$, e$^-$, HCO$^+$ \\
{ CO      +   He$^+$      $\rightarrow$                  He    +     O   +       C$^+$} &  He$^+$, N$^+$, NH$_3^+$ \\
{ H$_2$     +    CH$_3^+$      $\rightarrow$                 CH$_5^+$  +     Photon} & H$_2$CO, CH$_4$, CH$_5^+$ \\
{ HF    +     C$^+$      $\rightarrow$       H    +      CF$^+$}  & HF, CF$^+$ \\
{ Mg   +      HCO$^+$     $\rightarrow$      HCO    +    Mg$^+$} & Mg, Mg$^+$ \\
Na    +     HCO$^+$     $\rightarrow$         HCO    +    Na$^+$  & Na, Na$^+$ \\
{ CF$^+$  +      e$^-$      $\rightarrow$         C    +      F} & HF, CF$^+$ \\
{ HCS$^+$   +    e$^-$     $\rightarrow$         S     +     CH}  & HCS$^+$, OCS$^+$ \\
SiH$_2^+$ +     e$^-$      $\rightarrow$          H     +     SiH & SiH, SiN \\
HCNH$^+$ +     e$^-$      $\rightarrow$         H   +       H   +       CN & HNC, HCNH$^+$ \\
{ H$_3$O$^+$   +    e$^-$       $\rightarrow$    H    +      H   +       OH} & H$_2$O, H$_3$O$^+$ \\
GRAIN$^-$  +   Fe$^+$     $\rightarrow$      GRAIN0  +   Fe & Fe, Fe$^+$ \\
GRAIN$^-$   +  Mg$^+$     $\rightarrow$     GRAIN0   +  Mg & Mg, Mg$^+$ \\
GRAIN$^-$  +   Na$^+$       $\rightarrow$     GRAIN0   +  Na & Na, Na$^+$ \\
        \hline
    \end{tabular}
\\
$^*$ CR means cosmic-ray particle.\\
Boldface reactions are the reactions also listed in Table~\ref{key-reactions-1e5yr}.\\
    \label{key-reactions-1e6yr}
\end{table*}

\section{The new version of the Nautilus astrochemical code}

\subsection{Model presentation}

Because surface chemistry cannot be ignored, we have also created a complete full package for astrochemical simulations compatible with the kida.uva.2024 network, including all input gas-grain chemical files and a new version of the Nautilus fortran code. The Nautilus gas-grain code was presented in \citet{2016MNRAS.459.3756R} and was made public afterward. It has been used by many groups around the world. Briefly, Nautilus is a fortran code that computes chemical abundances (in the gas-phase and at the surface of interstellar grains) as a function of time for given physical conditions. The surface chemistry is based on the rate equation method \citep{1992ApJS...82..167H}. It is a three-phase code, meaning that a distinction is made for species at the surface of the grains between the few external monolayers of molecules (four layers in this version) and the rest of the mantle below. The species are more mobile at the surface than in the mantle. The mantle and surface phases are both affected by photo-dissociations involving direct and indirect UV photons. Species at the surface of the grains and in the mantles can diffuse by thermal hopping or through the tunneling effect (see below). Only  Langmuir-Hinshelwood diffusion is considered. Competition between diffusion, desorption, and reaction (as stated by \citet{2007A&A...469..973C} and \citet{2011ApJ...735...15G}) is also included in the model. The ratio between binding and diffusion energies (used to compute the diffusion energies from the binding energy) is set to 0.8 in the mantle and 0.4 at the surface following \citet{2016MNRAS.459.3756R}. In addition to thermal desorption, several other non-thermal desorption mechanisms are included: whole grain cosmic-ray heating, cosmic-ray sputtering (see below), photo-desorption, and chemical desorption (see below). Details on these processes can be found in \citet{2021A&A...652A..63W}. Only one grain size is considered (0.1 $\mu$m in radius) and the gas-to-dust mass ratio is set to 0.01. All the model parameters are explicitly given in the code and can be changed for specific applications. Documentation is provided with the code. 

As stated earlier, the kida.uva.2024 gas-phase chemical network is provided with a validity range of temperatures for each reaction. For one specific reaction, there might be several rate coefficients with complementary ranges of temperature. Depending on the gas temperature, Nautilus will select which rate coefficient to use. However, if the requested temperature lies outside the range defined in kida.uva.2024, Nautilus will not extrapolate the temperature dependent rates. The rate constant for temperatures above the upper (or lower) limit temperature will be extrapolated as a constant value equal to the rate at the  upper (or lower) limit temperature. If users want to change this, they need to modify the fortran code. 

Compared to the version from \citet{2016MNRAS.459.3756R}, Nautilus 2.0 has been upgraded in the following ways:
\begin{itemize}
    \item The code itself has been cleaned up, with some features that were either not used or not working being removed. As a consequence, the input parameter file {\it parameters.in} has been modified. 
    \item The activation energies for surface reactions are no longer read from the external file {\it activation-energies.in}. Instead, they are read from the surface reaction file in the third column for reaction type 14 (see Table \ref{tab:id_types}). 
    \item An additional way to compute the dust temperature was added using the prescription from \citet{2017A&A...604A..58H}, equation 8. This formula depends on the local visual extinction and UV field. 
    \item The formula for chemical desorption from \citet{2016A&A...585A..24M}, based on experiments, has been added. In addition, the code will switch automatically from rates for silicate surfaces to water covered grains when the number of monolayers exceeds four.
    \item In addition to thermal diffusion, all species are now allowed to diffuse by tunneling following the formalism of \citet{1992ApJS...82..167H} and depending on their mass. The barrier thickness is set to 2.5 angstrom \citep[see also][]{2014MNRAS.440.3557R}. 
    \item A new non-thermal desorption mechanism has been added to the model: sputtering by cosmic-rays (type 77 in Table \ref{tab:id_types}; see \citet{2021A&A...652A..63W} for details on the mechanism).
\end{itemize}
 Although the online KIDA database also stores some of the parameters for surface reactions (some binding and diffusion energies as well as chemical reactions), the online database is not complete, and most of the updates have been done over the years without proper documentation in the Nautilus-associated chemical networks. In addition, several processes (adsorption and thermal and non-thermal desorptions) are not included in KIDA. As such, we provide with Nautilus the chemical network and other appropriate chemical parameters for surface processes used in the current version of Nautilus compatible with kida.uva.2024 (meaning that the chemical species are the same in both networks). 
The code can be freely downloaded\footnote{ https://astrochem-tools.org/codes/}.

\subsection{Comparison with TMC-1 (CP) observed chemical composition}

\begin{figure}
\centering
\includegraphics[width=1\linewidth]{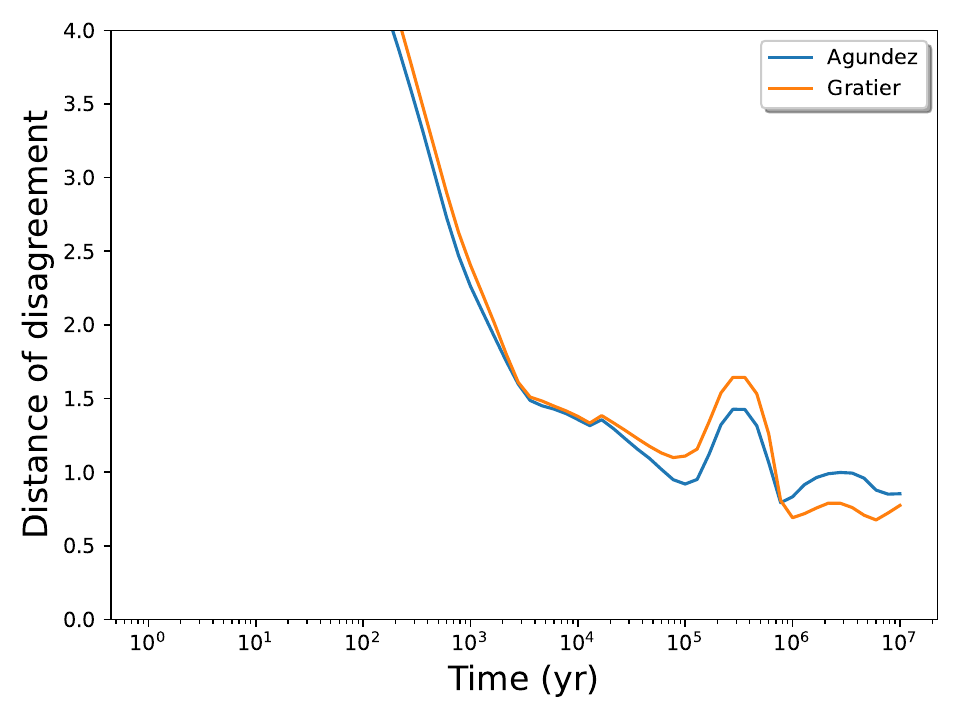}
\caption {Distance of disagreement between the modeled and observed abundances as a function of time. The two lines show the comparison for the two sets of observed abundances from  \citet{2013ChRv..113.8710A} and  \citet{2016ApJS..225...25G}.} \label{distance}
\end{figure}

\begin{figure*}
\centering
\includegraphics[width=1\linewidth]{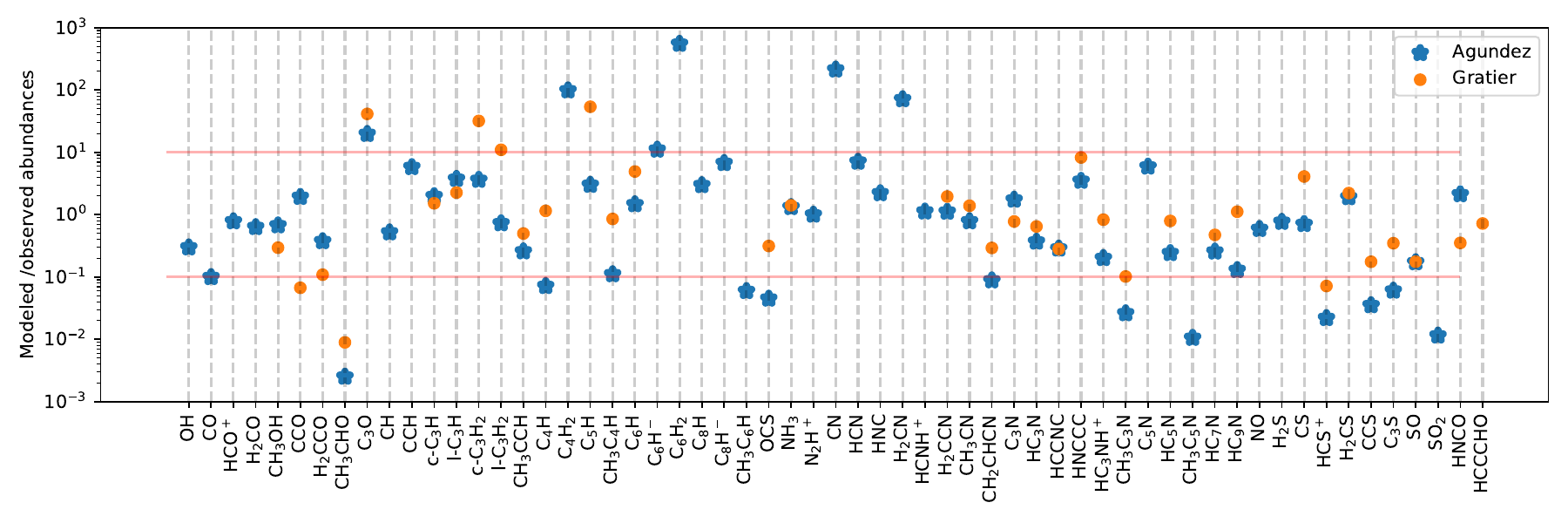}
\caption {Ratio between the modeled abundances at the "best times" and the observed abundances. The horizontal lines present ratios of ten and 0.1.   } \label{mod_vs_obs_both}
\end{figure*}

The final goal of astrochemical models is to reproduce the chemical composition observed in the ISM. To provide a reference, we present a comparison between the simulated time dependent chemical abundances of the new model with observations in the TMC-1 (CP) cold core, which is the most studied source of this type. Indeed, more than 50 molecules have been detected in this source, and their abundances are published in the literature. To compare with the models, we used two different sets of observed abundances. The first one has been gathered by \citet{2013ChRv..113.8710A} and contains 54 molecules published in the literature. We do not take into account the upper or lower limits. The second set of abundances has been obtained by \citet{2016ApJS..225...25G} by reanalyzing the publicly available observations from \citet{2004PASJ...56...69K}. It contains 37 molecules with corresponding abundances. Almost all species listed in \citeauthor{2016ApJS..225...25G} are also present in \citeauthor{2013ChRv..113.8710A}, except for one species (HCCCHO). We note that we did not include the recent detections with the YEBES telescope and the Quichote spectral survey \citep{refId0} because the reactions currently in KIDA do not allow us to describe the chemistry of these complex species. To compare the model results with the observations, we computed the distance of disagreement \citep[see for instance][]{2010SSRv..156...13W}. At each time step, the parameter D was computed according to
\begin{equation}
\rm D(t) = \frac{1}{N_i} \sum_i  \mid (log(X_{mod,i} (t)) -  log(X_{obs,i})) \mid,
\end{equation}
with $\rm N_i$ as the number of observed species, $\rm X_{mod,i} (t)$ as the modeled abundance at time t of species i, and $\rm X_{obs,i}$ as the observed abundance of species i. For the model, we used the physical conditions constrained by \citet{2019A&A...624A.105F} using the line observed excitation: a gas and grain temperature of 10~K and an H density of $3\times 10^4$~cm$^{-3}$. A standard cosmic-ray ionization rate of $10^{-17}$~s$^{-1}$ and a high visual extinction of 15 was assumed. The species are initially in atomic form (neutral or ionized) except for hydrogen, which is assumed to be entirely molecular. The elemental abundances are the same as in \citet[][their Table 1]{2017MNRAS.469..435V}. The gas-grain model was then run for $10^7$~yr. The computed distance of disagreement D as a function of time for both sets of observations is shown in Fig.~\ref{distance}. The agreement between the two sets of observations and the model is rather similar, with a slightly lower value of D when using the list of \citeauthor{2016ApJS..225...25G}, but the "best" time is different. The minimum value of D is 0.79 at  $7.7\times 10^5$ yr when comparing with \citet{2013ChRv..113.8710A}, and it is 0.67 at $6\times 10^6$~yr when comparing with \citet{2016ApJS..225...25G}. This means that, at the best time, the mean model abundances are within a factor (divided or multiplied) of six in the first case and a factor of five in the second when compared with the observations. The difference of agreement may, however, not be significant when considering the uncertainties in the observed abundances. We may conclude that both sets of abundances are well reproduced for times close to or larger than $10^6$~yr.
We note that this level of agreement is similar (although slightly better) to the one obtained with the previous version of kida.uva.2014 and of Nautilus \citep[shown in][]{2015ApJS..217...20W}. For the two "best times," we plot the ratio between the observed and modeled abundances in Fig.~\ref{mod_vs_obs_both}. Most species are reproduced within a factor of ten. For some species, the level of agreement depends on the choice of the observed value. For C$_5$H, for instance, the model reproduces the observed value from \citet{2013ChRv..113.8710A}, while it does not reproduce the one of \citet{2016ApJS..225...25G} . For OCS, we observed the contrary. Overall, 16 of the 57 species from the list of \citeauthor{2013ChRv..113.8710A} and six of the 37 species from the list of \citeauthor{2016ApJS..225...25G} list are not reproduced within a factor of ten. 

\section{Conclusions}

In this article, we have presented the new kida.uva.2024 gas-phase chemical network for ISM models. This is an update of the kida.uva.2011 \citep{2012ApJS..199...21W} and kida.uva.2014 \citep{2015ApJS..217...20W} networks. In addition, we have introduced the new version of the astrochemical gas-grain code Nautilus (with appropriate chemical networks for surface processes). The kida.uva.2024 chemical network comes with a number of warnings:
\begin{itemize}
\item Several reactions have been added to the kida.uva.2024 network to ensure a better description of their chemistry, although their rate coefficients and the products of these reactions are highly uncertain. In fact, this is also the case for many reactions inherited from the OSU network and for which we do not have any reference in the KIDA online database. 
\item Some rate coefficients are given in the network with temperature ranges of validity. The user needs to decide how to use them outside of their valid range (extrapolation or otherwise).
\item Many of the reactions included in the network still do not have any associated reference. Although we are correcting this over time for potentially wrong data, we cannot exclude the fact that mistakes may remain.
\item The network may not be complete for some species, in particular complex organic molecules.
\end{itemize}

In conclusion, we have provided the community with a full documented package to simulate the interstellar chemistry. The chemical files are formatted and can be used independently from Nautilus.


\begin{acknowledgements}
The maintenance and development of the KIDA database are possible thanks to the Observatoire Aquitain des Sciences de l’Univers and the Laboratoire d'astrophysique de Bordeaux. 
The KIDA team is also grateful to the persons who submitted data to the database or helped correct existing values since 2015: Daniel Seifried, Rafael M. Vichietti, Patrice Theulé, Sergio Rampino, Magnus Gustafsson, Tommaso Grassi, Jon Ramsey, Benoit Tabone, Bjoern Bastian, Alan Heays, Mathieu Bertin, Philipp Schmid, Olivia Venot, Daniela Ascenzi, Kevin Douglas, Venkatesan Thimmakondu Samy, Nicolas West, Karl-Mikael Svensson, Jessica Perrero, Ocana Antonio, James Babb,  Ioan Schneider, and Félix Iacob.
     
\end{acknowledgements}

\bibliographystyle{aa} 
\bibliography{bib} 

\begin{thebibliography}{190}
\expandafter\ifx\csname natexlab\endcsname\relax\def\natexlab#1{#1}\fi

\bibitem[{Adams \& Smith(1976)}]{Adams1976}
Adams, N.~G. \& Smith, D. 1976, Journal of Physics B Atomic Molecular Physics,
  9, 1439

\bibitem[{Adams {et~al.}(1984)Adams, Smith, \& Millar}]{Adams1984}
Adams, N.~G., Smith, D., \& Millar, T.~J. 1984, Monthly Notices of The Royal
  Astronomical Society, 211, 857

\bibitem[{Adusei {et~al.}(1996)Adusei, Blue, \& Fontijn}]{Adusei1996}
Adusei, G.~Y., Blue, A.~S., \& Fontijn, A. 1996, The Journal of Physical
  Chemistry, 100, 16921

\bibitem[{{Ag{\'u}ndez} \& {Wakelam}(2013)}]{2013ChRv..113.8710A}
{Ag{\'u}ndez}, M. \& {Wakelam}, V. 2013, Chemical Reviews, 113, 8710

\bibitem[{Amano(1990)}]{Amano1990}
Amano, T. 1990, Journal of Chemical Physics, 92, 6492

\bibitem[{Anicich(1993)}]{Anicich1993}
Anicich, V.~G. 1993, Journal of Physical and Chemical Reference Data, 22, 1469

\bibitem[{Anicich(2003)}]{Anicich2003}
Anicich, V.~G. 2003

\bibitem[{Anicich \& Huntress(1986)}]{Anicich1986}
Anicich, V.~G. \& Huntress, Jr., W.~T. 1986, Astrophysical Journal Supplement
  Series, 62, 553

\bibitem[{Antipov {et~al.}(2009)Antipov, Sjolander, Nyman, \&
  Gustafsson}]{Antipov2009}
Antipov, S.~V., Sjolander, T., Nyman, G., \& Gustafsson, M. 2009, Journal of
  Chemical Physics, 131, 074302

\bibitem[{Antiñolo {et~al.}(2016)Antiñolo, Agúndez, Jiménez, Ballesteros,
  Canosa, El~Dib, Albaladejo, \& Cernicharo}]{Antinolo2016}
Antiñolo, M., Agúndez, M., Jiménez, E., {et~al.} 2016, Astrophysical
  Journal, 823, 25

\bibitem[{Arthur \& Bell(1978)}]{Arthur1978}
Arthur, N.~L. \& Bell, T.~N. 1978, Rev. Chem. Intermed., 2, 37

\bibitem[{Atkinson {et~al.}(2004)Atkinson, Baulch, Cox, Crowley, Hampson,
  Hynes, Jenkin, Rossi, \& Troe}]{Atkinson2004}
Atkinson, R., Baulch, D.~L., Cox, R.~A., {et~al.} 2004, Atmospheric Chemistry
  \& Physics, 4, 1461

\bibitem[{Avramenko \& Krasnen'kov(1967)}]{Avramenko1967}
Avramenko, L.~I. \& Krasnen'kov, V.~M. 1967, Bull. Acad. Sci. USSR Div. Chem.
  Sci. (Engl. Transl.), 16, 501

\bibitem[{Azatyan {et~al.}(1975)Azatyan, Aleksandrov, \& Troshin}]{Azatyan1975}
Azatyan, V.~V., Aleksandrov, E.~N., \& Troshin, A.~F. 1975, Kinetics and
  Catalysis, 16, 261

\bibitem[{Barckholtz {et~al.}(2001)Barckholtz, Snow, \&
  Bierbaum}]{Barckholtz2001}
Barckholtz, C., Snow, T.~P., \& Bierbaum, V.~M. 2001, Astrophysical Journal
  Letters, 547, L171

\bibitem[{Bauerle {et~al.}(1995{\natexlab{a}})Bauerle, Klatt, \&
  Wagner}]{Bauerle1995a}
Bauerle, S., Klatt, M., \& Wagner, H.~G. 1995{\natexlab{a}}, Ber. Bunsenges.
  Phys. Chem., 99, 97

\bibitem[{Bauerle {et~al.}(1995{\natexlab{b}})Bauerle, Klatt, \&
  Wagner}]{Bauerle1995b}
Bauerle, S., Klatt, M., \& Wagner, H.~G. 1995{\natexlab{b}}, Berichte Der
  Bunsen-Gesellschaft-Physical Chemistry Chemical Physics, 99, 870

\bibitem[{Baulch {et~al.}(2005)Baulch, Bowman, Cobos, Cox, Just, Kerr, Pilling,
  Stocker, Troe, Tsang, Walker, \& Warnatz}]{Baulch2005}
Baulch, D.~L., Bowman, C.~T., Cobos, C.~J., {et~al.} 2005, J. Phys. Chem. Ref.
  Data, 34, 757

\bibitem[{Baulch {et~al.}(1994)Baulch, Cobos, Cox, Frank, Hayman, Just, Kerr,
  Murrells, Pilling, Troe, Walker, \& Warnatz}]{Baulch1994}
Baulch, D.~L., Cobos, C.~J., Cox, R.~A., {et~al.} 1994, Journal of Physical and
  Chemical Reference Data, 23, 847

\bibitem[{Becker {et~al.}(2000)Becker, Kurtenbach, Schmidt, \&
  Wiesen}]{Becker2000}
Becker, K.~H., Kurtenbach, R., Schmidt, F., \& Wiesen, P. 2000, Combust. Flame,
  120, 570

\bibitem[{Berteloite {et~al.}(2010)Berteloite, Le~Picard, Balucani, Canosa, \&
  Sims}]{Berteloite2010}
Berteloite, C., Le~Picard, S., Balucani, N., Canosa, A., \& Sims, I. 2010,
  Physical Chemistry Chemical Physics, 12, 3677

\bibitem[{Bettens \& Herbst(1995)}]{Bettens1995}
Bettens, R. P.~A. \& Herbst, E. 1995, International Journal of Mass
  Spectrometry and Ion Processes, 149, 321

\bibitem[{Biennier {et~al.}(2014)Biennier, Carles, Cordier, Guillemin, Picard,
  \& Faure}]{Biennier2014}
Biennier, L., Carles, S., Cordier, D., {et~al.} 2014, Icarus, 227, 123

\bibitem[{Bohland {et~al.}(1985)Bohland, Dobe, Temps, \& Wagner}]{Bohland1985}
Bohland, T., Dobe, S., Temps, F., \& Wagner, H.~G. 1985, Ber. Bunsenges. Phys.
  Chem., 89, 1110

\bibitem[{Bohme {et~al.}(1988)Bohme, Wlodek, \& Fox}]{Bohme1988}
Bohme, D.~K., Wlodek, S., \& Fox, A. 1988, {Chemical Pathways from Atomic
  Silicon Ions to Silicon Carbides and Oxides}, ed. T.~J. M. . D.~A. Williams
  (Kluwer Academic Publishers, Dordrecht, Boston)

\bibitem[{Brownsword {et~al.}(1997)Brownsword, Hancock, \&
  Heard}]{Brownsword1997}
Brownsword, R.~A., Hancock, G., \& Heard, D. 1997, J. Chem. Soc. Faraday
  Trans., 93, 2473

\bibitem[{Bryukov {et~al.}(2001)Bryukov, Slagle, \& Knyazev}]{Bryukov2001}
Bryukov, M.~G., Slagle, I.~R., \& Knyazev, V.~D. 2001, Journal of Physical
  Chemistry A, 105, 6900

\bibitem[{Canosa {et~al.}(1995)Canosa, Laub{\'e}, Rebrion, Pasquerault, Gomet,
  \& Rowe}]{Canosa1995}
Canosa, A., Laub{\'e}, S., Rebrion, C., {et~al.} 1995, Chemical Physics
  Letters, 245, 407

\bibitem[{Carles {et~al.}(2011)Carles, Adjali, Monnerie, Guillemin, \&
  Le~Garrec}]{Carles2011}
Carles, S., Adjali, F., Monnerie, C., Guillemin, J.-C., \& Le~Garrec, J.-L.
  2011, Icarus, 211, 901

\bibitem[{Carty {et~al.}(2006)Carty, Goddard, Kahler, Sims, \&
  Smith}]{Carty2006}
Carty, D., Goddard, A., Kahler, S., Sims, I., \& Smith, I. 2006, J. Phys. Chem.
  A, 110, 3101

\bibitem[{{Cazaux} {et~al.}(2005){Cazaux}, {Caselli}, {Tielens}, {LeBourlot},
  \& {Walmsley}}]{2005JPhCS...6..155C}
{Cazaux}, S., {Caselli}, P., {Tielens}, A.~G.~G.~M., {LeBourlot}, J., \&
  {Walmsley}, M. 2005, in Journal of Physics Conference Series, Vol.~6, Journal
  of Physics Conference Series, 155--160

\bibitem[{{Cernicharo, José} {et~al.}(2022){Cernicharo, José}, {Agúndez,
  Marcelino}, {Cabezas, Carlos}, {Marcelino, Nuria}, {Tercero, Belén}, {Pardo,
  Juan Ramón}, {Fuentetaja, Raúl}, \& {de Vicente, Pablo}}]{refId0}
{Cernicharo, José}, {Agúndez, Marcelino}, {Cabezas, Carlos}, {et~al.} 2022,
  EPJ Web of Conferences, 265, 00041

\bibitem[{Chabot {et~al.}(2013)Chabot, Beroff, Gratier, Jallat, \&
  Wakelam}]{Chabot2013}
Chabot, M., Beroff, K., Gratier, P., Jallat, A., \& Wakelam, V. 2013,
  Astrophysical Journal, 771, 90

\bibitem[{Chabot {et~al.}(2010)Chabot, Tuna, Beroff, Pino, Le~Padellec,
  Desequelles, Martinet, Nguyen-Thi, Carpentier, Le~Petit, Roueff, \&
  Wakelam}]{Chabot2010}
Chabot, M., Tuna, T., Beroff, K., {et~al.} 2010, Astronomy \& Astrophysics,
  524, A39

\bibitem[{{Chang} {et~al.}(2007){Chang}, {Cuppen}, \&
  {Herbst}}]{2007A&A...469..973C}
{Chang}, Q., {Cuppen}, H.~M., \& {Herbst}, E. 2007, \aap, 469, 973

\bibitem[{Cheikh Sid~Ely {et~al.}(2013)Cheikh Sid~Ely, Morales, Guillemin,
  Klippenstein, \& Sims}]{CheikhSidEly2013}
Cheikh Sid~Ely, S., Morales, S.~B., Guillemin, J.-C., Klippenstein, S.~J., \&
  Sims, I.~R. 2013, The Journal of Physical Chemistry A, 117, 12155

\bibitem[{Chenel {et~al.}(2010)Chenel, Mangaud, Justum, Talbi,
  Bacchus-Montabonel, \& Desouter-Lecomte}]{Chenel2010}
Chenel, A., Mangaud, E., Justum, Y., {et~al.} 2010, Journal of Physics B Atomic
  Molecular Physics, 43, 245701

\bibitem[{Cohen \& Westberg(1991)}]{Cohen1991}
Cohen, N. \& Westberg, K. 1991, J. Phys. Chem. Ref. Data, 20, 1211

\bibitem[{Cole {et~al.}(2012)Cole, Wehres, Yang, \& Thomsen}]{Cole2012}
Cole, C., Wehres, N., Yang, Z., \& Thomsen, D. 2012, The Astrophysical Journal
  Letters, 754, L5

\bibitem[{Davidson \& Hanson(1990)}]{Davidson1990}
Davidson, D. \& Hanson, R. 1990, International Journal of Chemical Kinetics,
  22, 843

\bibitem[{Derkatch {et~al.}(1999)Derkatch, Al-Khalili, Vikor, Neau, Shi,
  Danared, Ugglas, \& Larsson}]{Derkatch1999}
Derkatch, A.~M., Al-Khalili, A., Vikor, L., {et~al.} 1999, Journal of Physics B
  Atomic Molecular Physics, 32, 3391

\bibitem[{Diau {et~al.}(1995)Diau, Halbgewachs, Smith, \& Lin}]{Diau1995}
Diau, E., Halbgewachs, M., Smith, A., \& Lin, M. 1995, International Journal of
  Chemical Kinetics, 27, 867

\bibitem[{Dombrowsky \& Wagner(1992)}]{Dombrowsky1992}
Dombrowsky, C. \& Wagner, H.~G. 1992, Ber. Bunsenges. Phys. Chem., 96, 1048

\bibitem[{{Du}(2021)}]{2021RAA....21...77D}
{Du}, F. 2021, Research in Astronomy and Astrophysics, 21, 077

\bibitem[{Ehlerding {et~al.}(2004)Ehlerding, Hellberg, Thomas, Kalhori,
  Viggiano, Arnold, Larsson, \& Af~Ugglas}]{Ehlerding2004}
Ehlerding, A., Hellberg, F., Thomas, R., {et~al.} 2004, Physical Chemistry
  Chemical Physics (Incorporating Faraday Transactions), 6, 949

\bibitem[{Eichelberger {et~al.}(2007)Eichelberger, Snow, Barckholtz, \&
  Bierbaum}]{Eichelberger2007}
Eichelberger, B., Snow, T.~P., Barckholtz, C., \& Bierbaum, V.~M. 2007,
  Astrophysical Journal, 667, 1283

\bibitem[{Epée~Epée {et~al.}(2016)Epée~Epée, Mezei, Motapon, Pop, \&
  Schneider}]{EpeeEpee2016}
Epée~Epée, M.~D., Mezei, J.~Z., Motapon, O., Pop, N., \& Schneider, I.~F.
  2016, Monthly Notices of the Royal Astronomical Society, 455, 276

\bibitem[{Frank(1986)}]{Frank1986}
Frank, P. 1986, Proceedings of the 15th International Symposium on Rarefied Gas
  Dynamics, edited by, V. Boffi and C. Cercignani, 1986, 2, 422

\bibitem[{Freund \& Palmer(1977)}]{Freund1977}
Freund, H. \& Palmer, H. 1977, Int. J. Chem. Kinet., 9, 887

\bibitem[{{Fuente} {et~al.}(2019){Fuente}, {Navarro}, {Caselli}, {Gerin},
  {Kramer}, {Roueff}, {Alonso-Albi}, {Bachiller}, {Cazaux}, {Commercon},
  {Friesen}, {Garc{\'\i}a-Burillo}, {Giuliano}, {Goicoechea}, {Gratier},
  {Hacar}, {Jim{\'e}nez-Serra}, {Kirk}, {Lattanzi}, {Loison}, {Malinen},
  {Marcelino}, {Mart{\'\i}n-Dom{\'e}nech}, {Mu{\~n}oz-Caro}, {Pineda},
  {Tafalla}, {Tercero}, {Ward-Thompson}, {Trevi{\~n}o-Morales},
  {Rivi{\'e}re-Marichalar}, {Roncero}, {Vidal}, \&
  {Ballester}}]{2019A&A...624A.105F}
{Fuente}, A., {Navarro}, D.~G., {Caselli}, P., {et~al.} 2019, \aap, 624, A105

\bibitem[{Fujii {et~al.}(1987)Fujii, Kakuda, Takeishi, \& Miyama}]{Fujii1987}
Fujii, N., Kakuda, T., Takeishi, N., \& Miyama, H. 1987, J. Phys. Chem., 91,
  2144

\bibitem[{{Garrod} \& {Pauly}(2011)}]{2011ApJ...735...15G}
{Garrod}, R.~T. \& {Pauly}, T. 2011, \apj, 735, 15

\bibitem[{Geppert {et~al.}(2004)Geppert, Thomas, Hellberg, Ehlerding,
  Osterdahl, af~Ugglas, \& Larsson}]{Geppert2004}
Geppert, W.~D., Thomas, R., Hellberg, F., {et~al.} 2004, Phys. Chem. Chem.
  Phys., 6, 3415

\bibitem[{{Gratier} {et~al.}(2016){Gratier}, {Majumdar}, {Ohishi}, {Roueff},
  {Loison}, {Hickson}, \& {Wakelam}}]{2016ApJS..225...25G}
{Gratier}, P., {Majumdar}, L., {Ohishi}, M., {et~al.} 2016, \apjs, 225, 25

\bibitem[{Gronowski \& Kolos(2014)}]{Gronowski2014}
Gronowski, M. \& Kolos, R. 2014, Astrophysical Journal, 792, 89

\bibitem[{Gustafsson {et~al.}(2012)Gustafsson, Antipov, Franz, \&
  G.}]{Gustafsson2012}
Gustafsson, M., Antipov, S., Franz, J., \& G., N. 2012, Journal of Chemical
  Physics, 137, 104301

\bibitem[{Gustafsson {et~al.}(2014)Gustafsson, Monge-Palacios, \&
  Nyman}]{Gustafsson2014}
Gustafsson, M., Monge-Palacios, M., \& Nyman, G. 2014, Journal of Chemical
  Physics, 140, 184301

\bibitem[{Gustafsson \& Nyman(2015)}]{Gustafsson2015}
Gustafsson, M. \& Nyman, G. 2015, Monthly Notices of Royal Astronomical
  Society, 448, 2562

\bibitem[{Hamberg {et~al.}(2010)Hamberg, Osterdahl, Thomas, Zhaunerchyk,
  Vigren, Kaminska, Af~Ugglas, Kallberg, Simonsson, Paal, Larsson, \&
  Geppert}]{Hamberg2010}
Hamberg, M., Osterdahl, F., Thomas, R.~D., {et~al.} 2010, Astronomy and
  Astrophysics, 514, A83

\bibitem[{Hanson \& Salimian(1982)}]{Hanson1982}
Hanson, R.~K. \& Salimian, S. 1982, {Survey of Rate Constants in the N/H/O
  System}, ed. J.~S.-V. W.C.~Gardiner (Springer US)

\bibitem[{Harada \& Herbst(2008)}]{Harada2008}
Harada, N. \& Herbst, E. 2008, Astrophysical Journal, 685, 272

\bibitem[{{Harada} {et~al.}(2010){Harada}, {Herbst}, \&
  {Wakelam}}]{2010ApJ...721.1570H}
{Harada}, N., {Herbst}, E., \& {Wakelam}, V. 2010, \apj, 721, 1570

\bibitem[{Harada {et~al.}(2010)Harada, Herbst, \& Wakelam}]{Harada2010}
Harada, N., Herbst, E., \& Wakelam, V. 2010, Astrophysical Journal, 721, 1570

\bibitem[{{Hasegawa} {et~al.}(1992){Hasegawa}, {Herbst}, \&
  {Leung}}]{1992ApJS...82..167H}
{Hasegawa}, T.~I., {Herbst}, E., \& {Leung}, C.~M. 1992, \apjs, 82, 167

\bibitem[{Heays {et~al.}(2017)Heays, Bosman, \& van Dishoeck}]{Heays2017}
Heays, A.~N., Bosman, A.~D., \& van Dishoeck, E.~F. 2017, \aap, 602, A105

\bibitem[{Hemsworth {et~al.}(1974)Hemsworth, Payzant, Schiff, \&
  Bohme}]{Hemsworth1974}
Hemsworth, R., Payzant, J., Schiff, H., \& Bohme, D. 1974, Chemical Physics
  Letters, 26, 417

\bibitem[{{Henning} \& {Semenov}(2013)}]{2013ChRv..113.9016H}
{Henning}, T. \& {Semenov}, D. 2013, Chemical Reviews, 113, 9016

\bibitem[{Henshaw {et~al.}(1987)Henshaw, MacDonald, Stedman, \&
  Coombe}]{Henshaw1987}
Henshaw, T.~L., MacDonald, M.~A., Stedman, D.~H., \& Coombe, R.~D. 1987, The
  Journal of Physical Chemistry, 91, 2838

\bibitem[{Herbrechtsmeier(1973)}]{Herbrechtsmeier1973}
Herbrechtsmeier, P. 1973, {Reactions of O(3P) Atoms with Unsaturated
  C3-Hydrocarbons}, ed. F.~J. Weinberg (Academic Press, London)

\bibitem[{Herbst(1985)}]{Herbst1985}
Herbst, E. 1985, Astrophysical Journal, 291, 226

\bibitem[{Herbst(1987)}]{Herbst1987}
Herbst, E. 1987, Astrophysical Journal, 313, 867

\bibitem[{Herbst {et~al.}(1989{\natexlab{a}})Herbst, Millar, Wlodek, \&
  Bohme}]{Herbst1989a}
Herbst, E., Millar, T.~J., Wlodek, S., \& Bohme, D.~K. 1989{\natexlab{a}},
  Astronomy \& Astrophysics, 222, 205

\bibitem[{Herbst {et~al.}(1989{\natexlab{b}})Herbst, Smith, Adams, \&
  McIntosh}]{Herbst1989b}
Herbst, E., Smith, D., Adams, N.~G., \& McIntosh, B.~J. 1989{\natexlab{b}}, J.
  Chem. Soc., Faraday Trans. 2, 85, 1655

\bibitem[{Herbst {et~al.}(2000)Herbst, Terzieva, \& Talbi}]{Herbst2000}
Herbst, E., Terzieva, R., \& Talbi, D. 2000, Monthly Notices of The Royal
  Astronomical Society, 311, 869

\bibitem[{Hickson {et~al.}(2013)Hickson, Loison, \& Caubet}]{Hickson2013}
Hickson, K., Loison, J.-C., \& Caubet, P. 2013, J. Phys. Chem. Lett., 4, 2843

\bibitem[{{Hickson} {et~al.}(2015){Hickson}, {Loison}, {Bourgalais}, {Capron},
  {Le Picard}, {Goulay}, \& {Wakelam}}]{2015ApJ...812..107H}
{Hickson}, K.~M., {Loison}, J.-C., {Bourgalais}, J., {et~al.} 2015, \apj, 812,
  107

\bibitem[{Hickson {et~al.}(2015)Hickson, Loison, Bourgalais, Capron, Le~Picard,
  Goulay, \& Wakelam}]{Hickson2015}
Hickson, K.~M., Loison, J.-C., Bourgalais, J., {et~al.} 2015, Astrophysical
  Journal, 812, 107

\bibitem[{Hickson {et~al.}(2016)Hickson, Wakelam, \& Loison}]{Hickson2016}
Hickson, K.~M., Wakelam, V., \& Loison, J.-C. 2016, Molecular Astrophysics, 3,
  1

\bibitem[{{Hocuk} {et~al.}(2017){Hocuk}, {Sz{\H{u}}cs}, {Caselli}, {Cazaux},
  {Spaans}, \& {Esplugues}}]{2017A&A...604A..58H}
{Hocuk}, S., {Sz{\H{u}}cs}, L., {Caselli}, P., {et~al.} 2017, \aap, 604, A58

\bibitem[{{Holdship} {et~al.}(2017){Holdship}, {Viti}, {Jim{\'e}nez-Serra},
  {Makrymallis}, \& {Priestley}}]{2017AJ....154...38H}
{Holdship}, J., {Viti}, S., {Jim{\'e}nez-Serra}, I., {Makrymallis}, A., \&
  {Priestley}, F. 2017, \aj, 154, 38

\bibitem[{Hoyermann \& Nacke(1996)}]{Hoyermann1996}
Hoyermann, K. \& Nacke, F. 1996, Symp. Int. Combust. Proc., 26, 505

\bibitem[{Humpfer {et~al.}(1995)Humpfer, Oser, \& Grotheer}]{Humpfer1995}
Humpfer, R., Oser, H., \& Grotheer, H. 1995, International Journal of Chemical
  Kinetics, 27, 577

\bibitem[{Jensen {et~al.}(2000)Jensen, Bilodeau, Safvan, Seiersen, Andersen,
  Pedersen, \& Heber}]{Jensen2000}
Jensen, M.~J., Bilodeau, R.~C., Safvan, C.~P., {et~al.} 2000, Astrophysical
  Journal, 543, 764

\bibitem[{{Kaifu} {et~al.}(2004){Kaifu}, {Ohishi}, {Kawaguchi}, {Saito},
  {Yamamoto}, {Miyaji}, {Miyazawa}, {Ishikawa}, {Noumaru}, {Harasawa}, {Okuda},
  \& {Suzuki}}]{2004PASJ...56...69K}
{Kaifu}, N., {Ohishi}, M., {Kawaguchi}, K., {et~al.} 2004, \pasj, 56, 69

\bibitem[{Kalhori {et~al.}(2002)Kalhori, Viggiano, Arnold, Rosen, Semaniak,
  Derkatch, af~Ugglas, \& Larsson}]{Kalhori2002}
Kalhori, S., Viggiano, A.~A., Arnold, S.~T., {et~al.} 2002, Astronomy \&
  Astrophysics, 391, 1159

\bibitem[{Karkach \& Osherov(1999)}]{Karkach1999}
Karkach, S. \& Osherov, V. 1999, The Journal of Chemical Physics, 110, 11918

\bibitem[{Kathir {et~al.}(2017)Kathir, Nyman, \& Gustafsson}]{Kathir2017}
Kathir, R.~K., Nyman, G., \& Gustafsson, M. 2017, Monthly Notices of the Royal
  Astronomical Society, 470, 3068

\bibitem[{Kim \& Huntress~Jr.(1975)}]{Kim1975b}
Kim, J. \& Huntress~Jr., W. 1975, International Journal of Mass Spectrometry
  and Ion Physics, 16, 451

\bibitem[{Kim(1975)}]{Kim1975a}
Kim, J.~K. 1975, The Journal of Chemical Physics, 62, 2820

\bibitem[{{Le Bourlot} {et~al.}(2012){Le Bourlot}, {Le Petit}, {Pinto},
  {Roueff}, \& {Roy}}]{2012A&A...541A..76L}
{Le Bourlot}, J., {Le Petit}, F., {Pinto}, C., {Roueff}, E., \& {Roy}, F. 2012,
  \aap, 541, A76

\bibitem[{Le~Garrec {et~al.}(2003)Le~Garrec, Carles, Speck, Mitchell, Rowe, \&
  Ferguson}]{LeGarrec2003}
Le~Garrec, J.-L., Carles, S., Speck, T., {et~al.} 2003, Chemical Physics
  Letters, 372, 485

\bibitem[{Le~Garrec {et~al.}(1997)Le~Garrec, Lepage, Rowe, \&
  Ferguson}]{LeGarrec1997}
Le~Garrec, J.~L., Lepage, V., Rowe, B.~R., \& Ferguson, E.~E. 1997, Chemical
  Physics Letters, 270, 66

\bibitem[{{Le Teuff} {et~al.}(2000){Le Teuff}, {Millar}, \&
  {Markwick}}]{2000A&AS..146..157L}
{Le Teuff}, Y.~H., {Millar}, T.~J., \& {Markwick}, A.~J. 2000, \aaps, 146, 157

\bibitem[{Leen \& Graff(1988)}]{Leen1988}
Leen, T. \& Graff, M. 1988, Astrophysical Journal, 325, 411

\bibitem[{Lifshitz \& Michael(1991)}]{Lifshitz1991}
Lifshitz, A. \& Michael, J. 1991, Symposium (International) on Combustion, 23,
  59

\bibitem[{Lloyd(1974)}]{Lloyd1974}
Lloyd, A. 1974, Int. J. Chem. Kinet., 6, 169

\bibitem[{Loison {et~al.}(2016)Loison, Ag{\'undez}, Marcelino, Wakelam,
  Hickson, Cernicharo, Gerin, Roueff, \& Gu{\'elin}}]{Loison2016}
Loison, J.-C., Ag{\'undez}, M., Marcelino, N., {et~al.} 2016, \mnras, 456, 4101

\bibitem[{Loison {et~al.}(2017)Loison, Ag{\'undez}, Wakelam, Roueff, Gratier,
  Marcelino, Reyes, Cernicharo, \& Gerin}]{Loison2017}
Loison, J.-C., Ag{\'undez}, M., Wakelam, V., {et~al.} 2017, \mnras, 470, 4075

\bibitem[{Loison {et~al.}(2012)Loison, Halvick, Bergeat, Hickson, \&
  Wakelam}]{Loison2012}
Loison, J.-C., Halvick, P., Bergeat, A., Hickson, K.~M., \& Wakelam, V. 2012,
  Monthly Notices of The Royal Astronomical Society, 421, 1476

\bibitem[{Loison {et~al.}(2014{\natexlab{a}})Loison, Wakelam, \&
  Hickson}]{Loison2014a}
Loison, J.-C., Wakelam, V., \& Hickson, K. 2014{\natexlab{a}}, Monthly Notices
  of The Royal Astronomical Society, 443, 398

\bibitem[{Loison {et~al.}(2014{\natexlab{b}})Loison, Wakelam, Hickson, Bergeat,
  \& Mereau}]{Loison2014b}
Loison, J.-C., Wakelam, V., Hickson, K.~M., Bergeat, A., \& Mereau, R.
  2014{\natexlab{b}}, Monthly Notices of The Royal Astronomical Society, 437,
  930

\bibitem[{Luca {et~al.}(2002)Luca, Voulot, \& Gerlich}]{Luca2002}
Luca, A., Voulot, D., \& Gerlich, D. 2002, Proceedings of Contributed Papers,
  2, 204

\bibitem[{Mackay {et~al.}(1981)Mackay, Schiff, \& Bohme}]{Mackay1981}
Mackay, G., Schiff, H., \& Bohme, D. 1981, Canadian Journal of Chemistry, 59,
  1771

\bibitem[{Mackay {et~al.}(1980)Mackay, Vlachos, Bohme, \& Schiff}]{Mackay1980}
Mackay, G., Vlachos, G., Bohme, D., \& Schiff, H. 1980, International Journal
  of Mass Spectrometry and Ion Physics, 36, 259

\bibitem[{{Majumdar} {et~al.}(2018){Majumdar}, {Loison}, {Ruaud}, {Gratier},
  {Wakelam}, \& {Coutens}}]{Majumdar2017}
{Majumdar}, L., {Loison}, J.~C., {Ruaud}, M., {et~al.} 2018, \mnras, 473, L59

\bibitem[{Maluendes {et~al.}(1993)Maluendes, McLean, \& Herbst}]{Maluendes1993}
Maluendes, S.~A., McLean, A.~D., \& Herbst, E. 1993, Astrophysical Journal,
  417, 181

\bibitem[{Martinez {et~al.}(2008)Martinez, Betts, Villano, Eyet, Snow, \&
  Bierbaum}]{Martinez2008}
Martinez, Jr., O., Betts, N.~B., Villano, S.~M., {et~al.} 2008, Astrophysical
  Journal, 686, 1486

\bibitem[{Martinez {et~al.}(2010)Martinez, Yang, Demarais, Snow, \&
  Bierbaum}]{Martinez2010}
Martinez, Jr., O., Yang, Z., Demarais, N.~J., Snow, T.~P., \& Bierbaum, V.~M.
  2010, Astrophysical Journal, 720, 173

\bibitem[{Mayer \& Schieler(1967)}]{Mayer1967}
Mayer, S. \& Schieler, L.~Johnston, H. 1967, Symposium (International) on
  Combustion, 11, 837

\bibitem[{Mayer {et~al.}(1966)Mayer, Schieler, \& Johnston}]{Mayer1966}
Mayer, S., Schieler, L., \& Johnston, H. 1966, The Journal of Chemical Physics,
  45, 385

\bibitem[{{McElroy} {et~al.}(2013){McElroy}, {Walsh}, {Markwick}, {Cordiner},
  {Smith}, \& {Millar}}]{2013A&A...550A..36M}
{McElroy}, D., {Walsh}, C., {Markwick}, A.~J., {et~al.} 2013, \aap, 550, A36

\bibitem[{Mebel {et~al.}(1996)Mebel, Lin, Morokuma, \& Melius}]{Mebel1996}
Mebel, A., Lin, M., Morokuma, K., \& Melius, C. 1996, Int. J. Chem. Kinet., 28,
  693

\bibitem[{Millar {et~al.}(1987)Millar, Bennett, \& Herbst}]{Millar1987}
Millar, T.~J., Bennett, A., \& Herbst, E. 1987, Monthly Notices of The Royal
  Astronomical Society, 229, 41P

\bibitem[{{Millar} {et~al.}(1991){Millar}, {Bennett}, {Rawlings}, {Brown}, \&
  {Charnley}}]{1991A&AS...87..585M}
{Millar}, T.~J., {Bennett}, A., {Rawlings}, J.~M.~C., {Brown}, P.~D., \&
  {Charnley}, S.~B. 1991, \aaps, 87, 585

\bibitem[{{Millar} {et~al.}(1997){Millar}, {Farquhar}, \&
  {Willacy}}]{1997A&AS..121..139M}
{Millar}, T.~J., {Farquhar}, P.~R.~A., \& {Willacy}, K. 1997, \aaps, 121, 139

\bibitem[{Millar {et~al.}(1991)Millar, Herbst, \& Charnley}]{Millar1991}
Millar, T.~J., Herbst, E., \& Charnley, S. 1991, Astrophysical Journal, 369,
  147

\bibitem[{{Millar} {et~al.}(2024){Millar}, {Walsh}, {Van de Sande}, \&
  {Markwick}}]{2024A&A...682A.109M}
{Millar}, T.~J., {Walsh}, C., {Van de Sande}, M., \& {Markwick}, A.~J. 2024,
  \aap, 682, A109

\bibitem[{Miller \& Melius(1988)}]{Miller1988}
Miller, J.~A. \& Melius, C. 1988, Symposium (International) on Combustion, 21,
  919

\bibitem[{{Minissale} {et~al.}(2022){Minissale}, {Aikawa}, {Bergin}, {Bertin},
  {Brown}, {Cazaux}, {Charnley}, {Coutens}, {Cuppen}, {Guzman}, {Linnartz},
  {McCoustra}, {Rimola}, {Schrauwen}, {Toubin}, {Ugliengo}, {Watanabe},
  {Wakelam}, \& {Dulieu}}]{2022ESC.....6..597M}
{Minissale}, M., {Aikawa}, Y., {Bergin}, E., {et~al.} 2022, ACS Earth and Space
  Chemistry, 6, 597

\bibitem[{{Minissale} {et~al.}(2016){Minissale}, {Dulieu}, {Cazaux}, \&
  {Hocuk}}]{2016A&A...585A..24M}
{Minissale}, M., {Dulieu}, F., {Cazaux}, S., \& {Hocuk}, S. 2016, \aap, 585,
  A24

\bibitem[{Mitchell(1984{\natexlab{a}})}]{Mitchell1984b}
Mitchell, G.~F. 1984{\natexlab{a}}, Astrophysical Journal Supplement Series,
  54, 81

\bibitem[{Mitchell(1984{\natexlab{b}})}]{Mitchell1984a}
Mitchell, G.~F. 1984{\natexlab{b}}, Astrophysical Journal, 287, 665

\bibitem[{Miyoshi {et~al.}(1993)Miyoshi, Ohmori, Tsuchiya, \&
  Matsui}]{Miyoshi1993}
Miyoshi, A., Ohmori, K., Tsuchiya, K., \& Matsui, H. 1993, Chemical Physics
  Letters, 204, 241

\bibitem[{Montaigne {et~al.}(2005)Montaigne, Geppert, Semaniak, Osterdahl,
  Hellberg, Thomas, af~Ugglas, Roberts, Millar, Zhaunerchyk, Kaminska,
  Al-Khalili, Kallberg, \& Larsson}]{Montaigne2005}
Montaigne, H., Geppert, W.~D., Semaniak, J., {et~al.} 2005, Astrophysical
  Journal, 631, 653

\bibitem[{Morales {et~al.}(2010)Morales, Le~Picard, Canosa, \&
  Sims}]{Morales2010}
Morales, S.~B., Le~Picard, S.~D., Canosa, A., \& Sims, I.~R. 2010, Faraday
  Discussions, 147, 155

\bibitem[{Neufeld \& Wolfire(2009)}]{Neufeld2009}
Neufeld, D.~A. \& Wolfire, M.~G. 2009, Astrophysical Journal, 706, 1594

\bibitem[{Neufeld {et~al.}(2005)Neufeld, Wolfire, \& Schilke}]{Neufeld2005}
Neufeld, D.~A., Wolfire, M.~G., \& Schilke, P. 2005, Astrophysical Journal,
  628, 260

\bibitem[{Nguyen {et~al.}(2004)Nguyen, Zhang, Peeters, Truong, \&
  Nguyen}]{Nguyen2004}
Nguyen, H. M.~T., Zhang, S., Peeters, J., Truong, T.~N., \& Nguyen, M.~T. 2004,
  Chemical Physics Letters, 388, 94

\bibitem[{Oca{\~na} {et~al.}(2017)Oca{\~na}, Jim{\'enez}, Ballesteros, Canosa,
  Anti{\~nolo}, Albaladejo, Ag{\'undez}, Cernicharo, Zanchet, del Mazo,
  Roncero, \& Aguado}]{Ocana2017}
Oca{\~na}, A.~J., Jim{\'enez}, E., Ballesteros, B., {et~al.} 2017, \apj, 850,
  28

\bibitem[{Patterson \& Greene(1962)}]{Patterson1962}
Patterson, W. \& Greene, E. 1962, The Journal of Chemical Physics, 36, 1146

\bibitem[{Payzant {et~al.}(1975)Payzant, Schiff, \& Bohme}]{Payzant1975}
Payzant, J.~D., Schiff, H.~I., \& Bohme, D.~K. 1975, The Journal of Chemical
  Physics, 63, 149

\bibitem[{Peters \& Mahnen(1973)}]{Peters1973}
Peters, I. \& Mahnen, G. 1973, {Structure of Ethylene-Oxygen Flames. Reaction
  Mechanism and Rate Constants of Elementary Reactions}, ed. W.~F.J. (Academic
  Press, New York)

\bibitem[{Páramo {et~al.}(2006)Páramo, Canosa, Le~Picard, \&
  Sims}]{Paramo2006}
Páramo, A., Canosa, A., Le~Picard, S.~D., \& Sims, I.~R. 2006, The Journal of
  Physical Chemistry A, 110, 3121

\bibitem[{{Quan} {et~al.}(2008){Quan}, {Herbst}, {Millar}, {Hassel}, {Lin},
  {Guo}, {Honvault}, \& {Xie}}]{2008ApJ...681.1318Q}
{Quan}, D., {Herbst}, E., {Millar}, T.~J., {et~al.} 2008, \apj, 681, 1318

\bibitem[{Quan {et~al.}(2010)Quan, Herbst, Osamura, \& Roueff}]{Quan2010}
Quan, D., Herbst, E., Osamura, Y., \& Roueff, E. 2010, Astrophysical Journal,
  725, 2101

\bibitem[{R. {et~al.}(2008)R., Mikosch, Trippel, Weidemuller, \&
  R.}]{OttoR.2008}
R., O., Mikosch, J., Trippel, S., Weidemuller, M., \& R., W. 2008, Physical
  Review Letters, 101, 063201

\bibitem[{Rampino {et~al.}(2016)Rampino, Pastore, Garcia, Pacifici, \&
  LaganÃ }]{Rampino2016}
Rampino, S., Pastore, M., Garcia, E., Pacifici, L., \& LaganÃ , A. 2016,
  Monthly Notices of the Royal Astronomical Society, 460, 2368

\bibitem[{{Reboussin} {et~al.}(2014){Reboussin}, {Wakelam}, {Guilloteau}, \&
  {Hersant}}]{2014MNRAS.440.3557R}
{Reboussin}, L., {Wakelam}, V., {Guilloteau}, S., \& {Hersant}, F. 2014,
  \mnras, 440, 3557

\bibitem[{Roberge {et~al.}(1991)Roberge, Jones, Lepp, \&
  Dalgarno}]{Roberge1991}
Roberge, W.~G., Jones, D., Lepp, S., \& Dalgarno, A. 1991, Astrophysical
  Journal Supplement Series, 77, 287

\bibitem[{Rodgers(1996)}]{Rodgers1996}
Rodgers, A. 1996, Chemical Physics Letters, 253, 313

\bibitem[{Ruaud {et~al.}(2015)Ruaud, Loison, Hickson, Gratier, Hersant, \&
  Wakelam}]{Ruaud2015}
Ruaud, M., Loison, J., Hickson, K.~M., {et~al.} 2015, Monthly Notices of The
  Royal Astronomical Society, 447, 4004

\bibitem[{{Ruaud} {et~al.}(2016){Ruaud}, {Wakelam}, \&
  {Hersant}}]{2016MNRAS.459.3756R}
{Ruaud}, M., {Wakelam}, V., \& {Hersant}, F. 2016, \mnras, 459, 3756

\bibitem[{Schofield(1973)}]{Schofield1973}
Schofield, K. 1973, Journal of Physical and Chemical Reference Data, 2, 25

\bibitem[{Shin \& Beauchamp(1986)}]{Shin1986}
Shin, S. \& Beauchamp, J. 1986, Journal of Physical Chemistry, 90, 1507

\bibitem[{Sims {et~al.}(1994)Sims, Queffelec, Defrance, Rebrion-Rowe, Travers,
  Bocherel, Rowe, \& Smith}]{Sims1994}
Sims, I.~R., Queffelec, J.-L., Defrance, A., {et~al.} 1994, Journal of Chemical
  Physics, 100, 4229

\bibitem[{Singleton \& Cvetanovic(1988)}]{Singleton1988}
Singleton, D.~L. \& Cvetanovic, R.~J. 1988, Journal of Physical and Chemical
  Reference Data, 17, 1377

\bibitem[{Smith {et~al.}(1984)Smith, Adams, \& Ferguson}]{Smith1984}
Smith, D., Adams, N.~H., \& Ferguson, E.~E. 1984, International Journal of Mass
  Spectrometry and Ion Processes, 61, 15

\bibitem[{Smith {et~al.}(1992)Smith, Spanel, \& Mayhew}]{Smith1992}
Smith, D., Spanel, P., \& Mayhew, C. 1992, International Journal of Mass
  Spectrometry and Ion Processes, 117, 457

\bibitem[{Smith {et~al.}(1994)Smith, Spanel, \& Millar}]{Smith1994}
Smith, D., Spanel, P., \& Millar, T.~J. 1994, Monthly Notices of The Royal
  Astronomical Society, 266, 31

\bibitem[{Smith \& Adams(1981)}]{Smith1981}
Smith, I. \& Adams, N.~G. 1981, Monthly Notices of Royal Astronomical Society,
  197, 377

\bibitem[{Snow {et~al.}(2009)Snow, Stepanovic, Betts, Eichelberger, Martinez,
  \& Bierbaum}]{Snow2009}
Snow, T.~P., Stepanovic, M., Betts, N.~B., {et~al.} 2009, Astrobiology, 9, 1001

\bibitem[{Stancil {et~al.}(1993)Stancil, Babb, \& Dalgarno}]{Stancil1993}
Stancil, P.~C., Babb, J.~F., \& Dalgarno, A. 1993, \apj, 414, 672

\bibitem[{Stewart {et~al.}(1989)Stewart, Larson, \& Golden}]{Stewart1989}
Stewart, P.~H., Larson, C.~W., \& Golden, D.~M. 1989, Combust. Flame, 75, 25

\bibitem[{Svensson {et~al.}(2015)Svensson, Gustafsson, \& Nyman}]{Svensson2015}
Svensson, S. K.-M., Gustafsson, M., \& Nyman, G. 2015, Journal of Physical
  Chemistry A, 119, 12263

\bibitem[{Szab{\'o} \& Gustafsson(2019)}]{Szabo2019}
Szab{\'o}, P. \& Gustafsson, M. 2019, \mnras, 483, 3574

\bibitem[{Talbi {et~al.}(2000)Talbi, Le~Padellec, \& Mitchell}]{Talbi2000}
Talbi, D., Le~Padellec, A., \& Mitchell, J. B.~A. 2000, Journal of Physics B:
  Atomic, Molecular and Optical Physics, 33, 3631

\bibitem[{Talbi \& Smith(2009)}]{Talbi2009}
Talbi, D. \& Smith, I. W.~M. 2009, Physical Chemistry Chemical Physics, 11,
  8477

\bibitem[{Thorne {et~al.}(1983)Thorne, Anicich, \& Huntress}]{Thorne1983}
Thorne, L.~R., Anicich, V.~G., \& Huntress, W.~T. 1983, Chemical Physics
  Letters, 98, 162

\bibitem[{Thorne {et~al.}(1984)Thorne, Anicich, Prasad, \&
  Huntress}]{Thorne1984}
Thorne, L.~R., Anicich, V.~G., Prasad, S.~S., \& Huntress, Jr., W.~T. 1984,
  Astrophysical Journal, 280, 139

\bibitem[{{Tinacci} {et~al.}(2023){Tinacci}, {Ferrada-Chamorro}, {Ceccarelli},
  {Pantaleone}, {Ascenzi}, {Maranzana}, {Balucani}, \&
  {Ugliengo}}]{2023ApJS..266...38T}
{Tinacci}, L., {Ferrada-Chamorro}, S., {Ceccarelli}, C., {et~al.} 2023, \apjs,
  266, 38

\bibitem[{Tizniti {et~al.}(2014)Tizniti, Le~Picard, Lique, Berteloite, Canosa,
  Alexander, \& Sims}]{Tizniti2014}
Tizniti, M., Le~Picard, S.~D., Lique, F.~c., {et~al.} 2014, Nat Chem, 6, 141

\bibitem[{Tonolo {et~al.}(2020)Tonolo, Lupi, \& Puzzarini}]{Tonolo2020}
Tonolo, F., Lupi, J., \& Puzzarini, C.~Barone, V. 2020, The Astrophysical
  Journal, 900, id.85

\bibitem[{Tsang(1992)}]{Tsang1992}
Tsang, W. 1992, Journal of Physical and Chemical Reference Data, 21, 753

\bibitem[{Tsang \& Hampson(1986)}]{Tsang1986}
Tsang, W. \& Hampson, R.~F. 1986, Journal of Physical and Chemical Reference
  Data, 15, 1087

\bibitem[{Tsang \& Herron(1991)}]{Tsang1991}
Tsang, W. \& Herron, J.~T. 1991, Journal of Physical and Chemical Reference
  Data, 20, 609

\bibitem[{van Dishoeck(1988)}]{vanDishoeck1988}
van Dishoeck, E.~F. 1988, {Photodissociation and photoionization processes},
  ed. T.~J. M. . D.~A. Williams (Kluwer Academic Publishers)

\bibitem[{Vandooren {et~al.}(1994)Vandooren, Bian, \&
  Van~Tiggelen}]{Vandooren1994}
Vandooren, J., Bian, J., \& Van~Tiggelen, P. 1994, Combustion and Flame, 98,
  402

\bibitem[{{Vidal} {et~al.}(2017){Vidal}, {Loison}, {Jaziri}, {Ruaud},
  {Gratier}, \& {Wakelam}}]{2017MNRAS.469..435V}
{Vidal}, T. H.~G., {Loison}, J.-C., {Jaziri}, A.~Y., {et~al.} 2017, \mnras,
  469, 435

\bibitem[{Vidal {et~al.}(2017)Vidal, Loison, Jaziri, Ruaud, Gratier, \&
  Wakelam}]{Vidal2017}
Vidal, T.~H.~G., Loison, J.-C., Jaziri, A.~Y., {et~al.} 2017, MNRAS, 469, 435

\bibitem[{Viggiano {et~al.}(2005)Viggiano, Ehlerding, Hellberg, Thomas,
  Zhaunerchyk, Geppert, Montaigne, Larsson, Kaminska, \&
  Osterdahl}]{Viggiano2005}
Viggiano, A.~A., Ehlerding, A., Hellberg, F., {et~al.} 2005, Journal of
  Chemical Physics, 122, 226101

\bibitem[{{Wakelam} {et~al.}(2021){Wakelam}, {Dartois}, {Chabot}, {Spezzano},
  {Navarro-Almaida}, {Loison}, \& {Fuente}}]{2021A&A...652A..63W}
{Wakelam}, V., {Dartois}, E., {Chabot}, M., {et~al.} 2021, \aap, 652, A63

\bibitem[{{Wakelam} {et~al.}(2012){Wakelam}, {Herbst}, {Loison}, {Smith},
  {Chandrasekaran}, {Pavone}, {Adams}, {Bacchus-Montabonel}, {Bergeat},
  {B{\'e}roff}, {Bierbaum}, {Chabot}, {Dalgarno}, {van Dishoeck}, {Faure},
  {Geppert}, {Gerlich}, {Galli}, {H{\'e}brard}, {Hersant}, {Hickson},
  {Honvault}, {Klippenstein}, {Le Picard}, {Nyman}, {Pernot}, {Schlemmer},
  {Selsis}, {Sims}, {Talbi}, {Tennyson}, {Troe}, {Wester}, \&
  {Wiesenfeld}}]{2012ApJS..199...21W}
{Wakelam}, V., {Herbst}, E., {Loison}, J.~C., {et~al.} 2012, \apjs, 199, 21

\bibitem[{{Wakelam} {et~al.}(2015){Wakelam}, {Loison}, {Herbst}, {Pavone},
  {Bergeat}, {B{\'e}roff}, {Chabot}, {Faure}, {Galli}, {Geppert}, {Gerlich},
  {Gratier}, {Harada}, {Hickson}, {Honvault}, {Klippenstein}, {Le Picard},
  {Nyman}, {Ruaud}, {Schlemmer}, {Sims}, {Talbi}, {Tennyson}, \&
  {Wester}}]{2015ApJS..217...20W}
{Wakelam}, V., {Loison}, J.~C., {Herbst}, E., {et~al.} 2015, \apjs, 217, 20

\bibitem[{Wakelam {et~al.}(2009)Wakelam, Loison, Herbst, Talbi, Quan, \&
  Caralp}]{Wakelam2009}
Wakelam, V., Loison, J.-C., Herbst, E., {et~al.} 2009, Astronomy and
  Astrophysics, 495, 513

\bibitem[{Wakelam {et~al.}(2015)Wakelam, Loison, Hickson, \&
  Ruaud}]{Wakelam2015}
Wakelam, V., Loison, J.-C., Hickson, K.~M., \& Ruaud, M. 2015, \mnras, 453, L48

\bibitem[{{Wakelam} {et~al.}(2019){Wakelam}, {Ruaud}, {Gratier}, \&
  {Bonnell}}]{2019MNRAS.486.4198W}
{Wakelam}, V., {Ruaud}, M., {Gratier}, P., \& {Bonnell}, I.~A. 2019, \mnras,
  486, 4198

\bibitem[{{Wakelam} {et~al.}(2010){Wakelam}, {Smith}, {Herbst}, {Troe},
  {Geppert}, {Linnartz}, {{\"O}berg}, {Roueff}, {Ag{\'u}ndez}, {Pernot},
  {Cuppen}, {Loison}, \& {Talbi}}]{2010SSRv..156...13W}
{Wakelam}, V., {Smith}, I.~W.~M., {Herbst}, E., {et~al.} 2010, \ssr, 156, 13

\bibitem[{Wallington {et~al.}(1988)Wallington, Skewes, Siegl, Wu, \&
  Japar}]{Wallington1988}
Wallington, T.~J., Skewes, L.~M., Siegl, W.~O., Wu, C.~H., \& Japar, S.~M.
  1988, Int. J. Chem. Kinet., 20, 867

\bibitem[{Warnatz(1984)}]{Warnatz1984}
Warnatz, J. 1984, {Rate coefficients in the C/H/O system}, ed. J.~e.
  W.~C.~Gardiner (Springer-Verlag, New York)

\bibitem[{West {et~al.}(2019)West, Millar, Van~de Sande, Rutter, Blitz, Decin,
  \& Heard}]{West2019}
West, N.~A., Millar, T.~J., Van~de Sande, M., {et~al.} 2019, The Astrophysical
  Journal, 885, 134

\bibitem[{{Woodall} {et~al.}(2007){Woodall}, {Ag{\'u}ndez}, {Markwick-Kemper},
  \& {Millar}}]{2007A&A...466.1197W}
{Woodall}, J., {Ag{\'u}ndez}, M., {Markwick-Kemper}, A.~J., \& {Millar}, T.~J.
  2007, \aap, 466, 1197

\bibitem[{{Woon} \& {Herbst}(2009)}]{2009ApJS..185..273W}
{Woon}, D.~E. \& {Herbst}, E. 2009, \apjs, 185, 273

\bibitem[{Woon \& Herbst(2009)}]{Woon2009}
Woon, D.~E. \& Herbst, E. 2009, Astrophysical Journal Supplement Series, 185,
  273

\bibitem[{{Xu} {et~al.}(2007){Xu}, {Xie}, {Honvault}, {Lin}, \&
  {Guo}}]{2007JChPh.127b4304X}
{Xu}, C., {Xie}, D., {Honvault}, P., {Lin}, S.~Y., \& {Guo}, H. 2007, \jcp,
  127, 024304

\bibitem[{Yang {et~al.}(1993)Yang, Yu, Lin, \& Melius}]{Yang1993}
Yang, D., Yu, T., Lin, M.~C., \& Melius, C.~F. 1993, Chem. Phys., 177, 271

\bibitem[{Yang {et~al.}(2011)Yang, Cole, Martinez, Carpenter, Snow, \&
  Bierbaum}]{Yang2011}
Yang, Z., Cole, C.~A., Martinez, J.~O., {et~al.} 2011, Astrophysical Journal,
  739, 19

\bibitem[{Zabarnick {et~al.}(1989)Zabarnick, Fleming, \& Lin}]{Zabarnick1989}
Zabarnick, S., Fleming, J., \& Lin, M. 1989, Int. J. Chem. Kinet., 21, 765

\bibitem[{Zanchet {et~al.}(2009)Zanchet, Bussery-Honvault, Jorfi, \&
  Honvault}]{Zanchet2009}
Zanchet, A., Bussery-Honvault, B., Jorfi, M., \& Honvault, P. 2009, Physical
  Chemistry Chemical Physics, 11, 6182–6191

\bibitem[{Zellner \& Ewig(1988)}]{Zellner1988}
Zellner, R. \& Ewig, F. 1988, J. Phys. Chem., 92, 2971

\bibitem[{Öström {et~al.}(2016)Öström, Bezrukov, Nyman, \&
  Gustafsson}]{Ostrom2016}
Öström, J., Bezrukov, D.~S., Nyman, G., \& Gustafsson, M. 2016, The Journal
  of Chemical Physics, 144

\end{thebibliography}

\begin{appendix} 
\section{Modifications with respect to the online KIDA database}\label{modifs-onlineKIDA}

\begin{table*}[ht]
    \centering
        \caption{Species from the online KIDA database not included in the kida.uva.2024 network.}
    \begin{tabular}{c}
        \hline
        \hline
        Species names\\
        \hline
N($^2$D), O($^1$D), (13)C, C($^1$D), SCl,  FO, Cl$_2$, (13)CO, FeH, $^1$CH$_2$, CN$_2$, N$_2$H, FO$_2$, \\
FNO, ClOH, OClO, ClO$_2$, MgH$_2$, HC$_2$N, N$_2$H$_2$, NO$_3$, SO$_3$, H$_2$SO, HC$_2$N$_2$,  N$_2$H$_3$, HNCSH, \\
HNCHS, HCOCN, CH$_3$NH, H$_2$C$_3$N, N$_2$H$_4$, CH$_2$CHO, C$_5$S,  HNCHSH, C$_3$H$_2$CN, NH$_2$CHSH, C$_2$H$_3$NH$_2$, \\
C$_4$H$_2$CN, CH$_3$COCN, C$_4$H$_3$CN, C$_4$H$_5$, C$_3$H$_5$CN, C$_5$H$_5$, C$_4$H$_7$, C$_5$H6, c-C6H$_5$, C$_4$H8, C$_5$H$_7$, C$_7$H$_5$, \\
C$_4$H$_9$, C$_5$H$_8$, C$_6$H$_7$, C$_7$H$_6$, C$_4$H$_{10}$, C$_5$H$_9$, C$_6$H$_8$, C$_7$H$_7$, C$_8$H$_6$, C$_5$H$_{10}$, C$_6$H$_9$, C$_7$H$_8$, C$_5$H$_{11}$, \\
C$_6$H$_{10}$, C$_7$H$_9$, C$_5$H$_{12}$, C$_6$H$_{11}$, C$_7$H$_{10}$, C$_6$H$_{12}$, C$_6$H$_{13}$, C$_6$H$_{14}$, C$_{14}$H$_{10}$, H$_2^+$(J=1 to 10), NH$_2^-$, \\
NCS$^+$, NCSH$^+$, CH$_2$NH$^+$, C$_4$N$_2^+$, NH$_2$CHS$^+$, C$_4$N$_2$H$^+$,   C$_2$H$_5$N$^+$, NH$_2$CH$_2$SH$^+$, C$_3$H$_6^+$, C$_3$H$_5$N$^+$, \\
C$_6$N$_2$H$^+$, C$_4$H$_6^+$, C$_3$H$_5$NH$^+$, C$_4$H$_5$N$^+$, C$_3$H$_8^+$, C$_4$H$_5$NH$^+$, C$_5$H$_5$N$^+$, C$_4$H$_8^+$, C$_5$H$_7^+$, C$_4$H$_9^+$, C$_5$H$_9^+$, \\
C$_7$H$_7^+$, C$_8$H$_6^+$, C$_6$H$_7$N$^+$, C$_8$H$_7^+$, C$_6$H$_7$NH$^+$, C$_9$H$_7^+$, C$_9$H$_8^+$, C$_{14}$H$_{10}^+$, 
C$_{15}$H$_9^+$\\
        \hline
    \end{tabular}
    \label{removed-from-KIDA}
\end{table*}

\begin{table*}[ht]
    \centering
        \caption{Species not in the online KIDA database but included in the kida.uva.2024 network.}
    \begin{tabular}{c}
        \hline
        \hline
        Species names\\
        \hline
GRAIN0, GRAIN$^-$, XH, H$_5$C$_3$O$^+$, CS$_2$, CS$_2^+$, HCS$_2^+$, H$_2$S$_3$, S$_3$, S$_4$, S$_5$, S$_6$, S$_7$, S$_8$, CH$_3$SH$^+$, \\
NHCHO, HOCHN, NH$_2$CO, NH$_2$CH$_2$O, NH$_2$CH$_2$OH, CNCN, NC$_3$N, FeH, MgH$_2$\\
        \hline
    \end{tabular}
    \label{additions-from-KIDA}
\end{table*}

Table~\ref{removed-from-KIDA} lists the species that are present in the KIDA online database and that were not included in the kida.uva.2024 network. Note that N($^2$D), O($^1$D), and $^1$CH2 refer to excited species while (13)C and (13)CO refer to isotopic species. H$_2^+$(J=1 to 10) are the excited rotational states of H$_2$. Table~\ref{additions-from-KIDA} list the species that are not present in the KIDA online database but are included in the  kida.uva.2024 network. Note that although GRAIN0, GRAIN-, and XH are not in KIDA because they are not real chemical species, they were already present in kida.uva.2011 and kida.uva.2015. GRAIN0 and GRAIN- represent neutral and negatively charged grains related to type 0 reactions (Table~\ref{tab:id_types}). The XH species is related to the case where H$_2$ formation is forced \citep[reaction type 10 and 11, as explained in section 3.2 of][]{2012ApJS..199...21W}. Users may want to use it in the case where they do not compute surface chemistry or in the case where the Langmuir-Hinshelwood surface mechanism does not work \citep[at high dust temperature]{2012A&A...541A..76L,2005JPhCS...6..155C}. Table~\ref{tab:added-reactions} lists the destruction reactions with cosmic-ray induced photons (CRP), direct UV photons (Photons), and with electrons added to kida.uva.2024. The three last columns of the Table give the parameters used to compute the  rate coefficients depending on the type of reactions \citep[Formula 1, 5, and 7 from][for reactions with electrons, Photons, and CRP respectively]{2012ApJS..199...21W}.


\clearpage
\onecolumn

 \begin{longtable}{llll}
     \caption{\label{tab:added-reactions} 
     Missing destruction reactions added to kida.uva.2024.
     }\\
\hline\hline
Reaction   & $\alpha$ & $\beta$ & $\gamma$ \\
\hline
\endfirsthead
\caption{Continued.}\\
\hline\hline
Reaction   & $\alpha$ & $\beta$ & $\gamma$ \\
\hline
\endhead
\hline
\endfoot
HSO  +  CRP $\rightarrow$    S   + OH & 1.500e+03 & 0.000e+00 & 0.000e+00  \\
HNCS   +    CRP   $\rightarrow$ NH    +     CS & 7.500e+02 & 0.000e+00 & 0.000e+00  \\
HNCS    +   CRP   $\rightarrow$ HNC  +      S   & 7.500e+02  & 0.000e+00 & 0.000e+00  \\
NCSH     +  CRP   $\rightarrow$ SH + CN & 6.000e+03 & 0.000e+00  & 0.000e+00  \\
HCCS   +    CRP  $\rightarrow$ CS + CH & 1.500e+03 & 0.000e+00 & 0.000e+00  \\
HCNS    +   CRP   $\rightarrow$ HCN + S & 2.500e+03 &  0.000e+00  & 0.000e+00  \\
HCNS   +    CRP  $\rightarrow$ CH + NS & 5.000e+02 & 0.000e+00  & 0.000e+00  \\
HSCN  +     CRP   $\rightarrow$ HS + CN & 1.500e+03 & 0.000e+00  & 0.000e+00  \\
HSCN   +    CRP    $\rightarrow$ HCN + S & 1.500e+03 & 0.000e+00  & 0.000e+00  \\
HC$_2$O  +     CRP   $\rightarrow$ CH + CO & 6.000e+03 & 0.000e+00  & 0.000e+00  \\
CH$_2$OH  +    CRP   $\rightarrow$ CH$_2$ + OH & 6.000e+03 & 0.000e+00  & 0.000e+00  \\
CH$_3$O  +     CRP   $\rightarrow$ CH$_3$ + O & 1.500e+03 & 0.000e+00  & 0.000e+00  \\
CH$_3$S  +     CRP   $\rightarrow$ CH$_3$ + S & 1.500e+03 & 0.000e+00 & 0.000e+00  \\
CH$_2$SH +     CRP   $\rightarrow$ CH$_2$ + HS & 1.500e+03 & 0.000e+00  & 0.000e+00  \\
H$_2$CCS$^ +$     CRP  $\rightarrow$ CS + CH$_2$ & 9.150e+02 & 0.000e+00 & 0.000e+00  \\
H$_2$CCS +     CRP  $\rightarrow$ H$_2$C$_2$S$^+$ + e$^-$ & 1.220e+03 & 0.000e+00 & 0.000e+00  \\
HC$_3$S +      CRP  $\rightarrow$ CS + CCH & 1.500e+03 & 0.000e+00  & 0.000e+00  \\
HC$_3$O  +     CRP  $\rightarrow$ CO + CCH & 1.500e+03 & 0.000e+00  & 0.000e+00  \\
C$_5$O  +      CRP   $\rightarrow$ CO + C$_3$ & 6.600e+03 & 0.000e+00  & 0.000e+00  \\
CH$_2$NH$_2$  +   CRP  $\rightarrow$ H$_2$ + H$_2$CN & 4.980e+03 & 0.000e+00  & 0.000e+00  \\
HCCCHS  +   CRP  $\rightarrow$ CH$_3$ + HS & 1.500e+03 & 0.000e+00 & 0.000e+00  \\
NH$_2$CHS  +   CRP   $\rightarrow$ HCS + NH$_2$ & 3.000e+03 & 0.000e+00 & 0.000e+00  \\
HCCCHO  +   CRP    $\rightarrow$ CO + C$_2$H$_2$ & 1.500e+03 & 0.000e+00 & 0.000e+00  \\
c-C$_3$H$_2$O +   CRP  $\rightarrow$ CO + C$_2$H$_2$ & 9.150e+02 & 0.000e+00 & 0.000e+00  \\
H$_2$C$_3$O +     CRP $\rightarrow$ CO + C$_2$H$_2$ & 1.500e+03 & 0.000e+00 & 0.000e+00  \\
CH$_2$CCH$_2$ +   CRP $\rightarrow$ CH$_2$ + C$_2$H$_2$ & 4.500e+02 & 0.000e+00 & 0.000e+00  \\
C$_6$N  +      CRP $\rightarrow$ CN + C$_5$ & 1.000e+03 & 0.000e+00 & 0.000e+00  \\
CH$_3$CHS +    CRP $\rightarrow$ CH$_2$ + H$_2$CS & 3.170e+03 & 0.000e+00 & 0.000e+00  \\
C$_3$H$_5$ +      CRP  $\rightarrow$ H + CH$_3$CCH & 5.000e+03 & 0.000e+00 & 0.000e+00  \\
C$_7$O +       CRP  $\rightarrow$ CO + C$_6$ & 6.600e+03 & 0.000e+00 & 0.000e+00  \\
CH$_3$OCH$_2$ +   CRP  $\rightarrow$ H$_2$CO + CH$_2$ & 1.720e+03 & 0.000e+00  & 0.000e+00  \\
CH$_3$OCH$_2$ +   CRP  $\rightarrow$ CH$_3$CHOH$^+$ +  e$^-$ & 1.120e+03 & 0.000e+00 & 0.000e+00  \\
NH$_2$CH$_2$SH +  CRP $\rightarrow$ CH$_2$SH + NH$_2$ & 1.242e+03 & 0.000e+00 & 0.000e+00  \\
NH$_2$CH$_2$SH +  CRP $\rightarrow$ CH$_2$NH$_2$ + HS & 1.242e+03 & 0.000e+00 & 0.000e+00  \\
NH$_2$CH$_2$SH +  CRP $\rightarrow$ NH$_2$CH$_2$S + H & 1.242e+03 & 0.000e+00  & 0.000e+00  \\
NH$_2$CH$_2$SH +  CRP $\rightarrow$ CH$_2$NH + H$_2$S & 1.242e+03 & 0.000e+00 & 0.000e+00  \\
NH$_2$CH$_2$SH +  CRP $\rightarrow$ H$_2$CS + NH$_3$ & 1.242e+03 & 0.000e+00 & 0.000e+00  \\
C$_8$N +       CRP  $\rightarrow$ CN + C$_7$ & 1.750e+03 & 0.000e+00 & 0.000e+00  \\
C$_3$H$_7$ +      CRP $\rightarrow$ H + CH$_3$CHCH$_2$ & 5.000e+03 & 0.000e+00 & 0.000e+00  \\
C$_9$O +       CRP $\rightarrow$ CO + C$_8$ & 6.600e+03 & 0.000e+00  & 0.000e+00  \\
C$_{10}$N +      CRP  $\rightarrow$ CN + C$_9$ & 1.750e+03 & 0.000e+00 & 0.000e+00  \\
C$_3$H$_8^+$      CRP $\rightarrow$ H + C$_3$H$_7$ & 5.000e+03 & 0.000e+00  & 0.000e+00  \\
H$_2$C$_3$S +     CRP $\rightarrow$ CS + C$_2$H$_2$ & 6.000e+03 & 0.000e+00  & 0.000e+00 \\
HSO    +    Photon $\rightarrow$ S + OH & 3.300e-10 & 0.000e+00 & 1.400e+00  \\
HSO   +     Photon $\rightarrow$ HSO$^+$ + e$^-$ & 6.200e-12 & 0.000e+00 & 3.100e+00   \\
HCCP  +     Photon $\rightarrow$ H + CCP & 1.900e-09 & 0.000e+00  & 1.700e+00  \\
H$_2$CN   +    Photon $\rightarrow$ H + HCN & 1.600e-09 & 0.000e+00 & 2.690e+00  \\
HNCS   +    Photon $\rightarrow$ NH + CS & 1.000e-09 & 0.000e+00 & 2.500e+00  \\
HNCS    +   Photon  $\rightarrow$ HNC + S & 1.000e-09 & 0.000e+00 & 2.500e+00  \\
NCSH    +   Photon  $\rightarrow$ CN + SH & 1.000e-09 & 0.000e+00 & 2.500e+00  \\
HCCS    +   Photon  $\rightarrow$ CS + CH & 1.900e-09 & 0.000e+00 & 1.700e+00  \\
HCNS    +   Photon $\rightarrow$ HCN + S & 1.600e-09 & 0.000e+00 & 2.500e+00  \\
HCNS    +   Photon  $\rightarrow$ CH + NS & 4.000e-10 & 0.000e+00 & 2.500e+00  \\
HSCN    +   Photon  $\rightarrow$ HS + CN & 1.800e-09 & 0.000e+00 & 2.500e+00  \\
HSCN    +   Photon $\rightarrow$ HCN + S & 2.000e-10 & 0.000e+00 & 2.500e+00  \\
HC$_2$O   +    Photon  $\rightarrow$ CO + CH & 1.900e-09 & 0.000e+00 & 1.700e+00  \\
NH$_2$CN  +    Photon $\rightarrow$ CN + NH$_2$ & 1.900e-09 & 0.000e+00 & 1.700e+00  \\
CH$_2$OH   +   Photon $\rightarrow$ CH$_2$ + OH & 1.900e-09 & 0.000e+00 & 1.700e+00  \\
CH$_3$O    +   Photon $\rightarrow$ CH$_3$ + O & 1.900e-09 & 0.000e+00 & 1.700e+00  \\
CH$_3$S    +   Photon $\rightarrow$ CH$_3$ + S & 1.000E-09 & 0.000E+00 & 1.000E+01 \\
CH$_2$SH   +   Photon $\rightarrow$ CH$_2$ + HS & 1.000E-09 & 0.000E+00  & 1.000E+01  \\
H$_2$CCS   +   Photon $\rightarrow$ CS + CH$_2$ & 1.400e-09 & 0.000e+00  & 1.580e+00  \\
H$_2$CCS   +   Photon $\rightarrow$ H$_2$C$_2$S$^+$ + e$^-$ & 6.600e-10 & 0.000e+00 & 2.010e+00  \\
HC$_3$S    +   Photon $\rightarrow$ CS + CCH & 1.900e-09 & 0.000e+00 & 1.700e+00  \\
HC$_3$O    +   Photon $\rightarrow$ CO + CCH & 1.900e-09 & 0.000e+00 & 1.700e+00  \\
C$_5$O     +   Photon $\rightarrow$ CO + C$_3$ & 7.000e-09 & 0.000e+00 & 1.580e+00  \\
CH$_2$NH$_2$  +   Photon $\rightarrow$ H$_2$ + H$_2$CN & 3.500e-09 & 0.000e+00 & 1.630e+00  \\
HCCCHS   +  Photon $\rightarrow$ CH$_3$ + HS & 3.500e-09 & 0.000e+00 & 1.630e+00  \\
NH$_2$CHS   +  Photon  $\rightarrow$ NH$_2$ + CS + H & 2.060e-09 & 0.000e+00 & 2.500e+00  \\
HCCCHO   +  Photon $\rightarrow$ CO + C$_2$H$_2$ & 1.900e-09 & 0.000e+00 & 1.700e+00  \\
c-C$_3$H$_2$O  +  Photon  $\rightarrow$ CO + C$_2$H$_2$ & 1.400e-09 & 0.000e+00 & 1.580e+00  \\
H$_2$C$_3$O   +   Photon  $\rightarrow$ CO + C$_2$H$_2$ & 1.900e-09 & 0.000e+00 & 1.700e+00  \\
CH$_2$CCH$_2$  +  Photon  $\rightarrow$ CH$_2$ + C$_2$H$_2$ & 2.550e-09 & 0.000e+00 & 1.700e+00  \\
C$_6$N   +     Photon  $\rightarrow$ CN + C$_5$ & 5.000e-10 & 0.000e+00 & 1.700e+00  \\
CH$_3$CHS  +   Photon $\rightarrow$ CH$_2$ + H$_2$CS & 1.400e-09 & 0.000e+00 & 2.280e+00  \\
C$_3$H$_5$   +    Photon $\rightarrow$ H + CH$_3$CCH & 1.000e-09 & 0.000e+00 & 1.700e+00  \\
C$_7$O     +   Photon $\rightarrow$ CO + C$_6$ & 7.000e-09 & 0.000e+00 & 1.580e+00  \\
CH$_3$OCH$_2$  +  Photon $\rightarrow$ H$_2$CO + CH$_2$ & 1.500e-09 & 0.000e+00 & 1.600e+00  \\
CH$_3$OCH$_2$   + Photon $\rightarrow$ CH$_3$CHOH$^+$ + e$^-$ & 4.600e-10 & 0.000e+00 & 2.280e+00  \\
NH$_2$CH$_2$SH  + Photon $\rightarrow$ CH$_2$SH + NH$_2$ & 6.619e-10 & 0.000e+00 & 1.580e+00  \\
NH$_2$CH$_2$SH  + Photon  $\rightarrow$ CH$_2$NH$_2$ + HS & 6.619e-10 & 0.000e+00 & 1.580e+00  \\
NH$_2$CH$_2$SH  + Photon $\rightarrow$ NH$_2$CH$_2$S + H & 6.619e-10 & 0.000e+00 & 1.580e+00  \\
NH$_2$CH$_2$SH  + Photon $\rightarrow$ CH$_2$NH + H$_2$S & 6.619e-10 & 0.000e+00 & 1.580e+00  \\
NH$_2$CH$_2$SH  + Photon $\rightarrow$ H$_2$CS + NH$_3$ & 6.619e-10 & 0.000e+00 & 1.580e+00  \\
C$_8$N   +     Photon $\rightarrow$ C$_8$N$^+$ + e$^-$ & 2.000e-10 & 0.000e+00 & 2.500e+00  \\
C$_8$N    +    Photon  $\rightarrow$ CN + C$_7$ & 5.000e-10 & 0.000e+00 & 1.700e+00  \\
C$_3$H$_7$    +   Photon $\rightarrow$ H + CH$_3$CHCH$_2$ & 1.000e-09 & 0.000e+00 & 1.700e+00  \\
C$_9$O      +  Photon  $\rightarrow$ CO + C$_8$ & 7.000e-09 & 0.000e+00 & 1.580e+00  \\
C$_{10}$N  +     Photon  $\rightarrow$ C$_{10}$N$^+$ + e$^-$ & 2.000e-10 & 0.000e+00 & 2.500e+00  \\
C$_{10}$N   +    Photon $\rightarrow$ CN + C$_9$ & 5.000e-10 & 0.000e+00 &  1.700e+00  \\
C$_3$H$_8$    +   Photon $\rightarrow$ H + C$_3$H$_7$ & 1.000e-09 & 0.000e+00 & 1.700e+00  \\
H$_2$C$_3$S    +  Photon $\rightarrow$ CS + C$_2$H$_2$ & 1.500e-09 & 0.000e+00  & 2.500e+00 \\
P$^+$        + e$^-$ $\rightarrow$ P + Photon & 4.000e-12 & -6.000e-01 & 0.000e+00 \\
CNC$^+$  +    e$^-$ $\rightarrow$ C + CN & 3.800e-07 & -6.000e-01 & 0.000e+00  \\
C$_2$N$^+$   +    e$^-$  $\rightarrow$ C + CN & 2.000e-08 & -6.000e-01  & 0.000e+00  \\
C$_2$N$^+$    +   e$^-$ $\rightarrow$ N + C$_2$ & 3.800e-07 & -6.000e-01 & 0.000e+00  \\
HSiO$^+$    +  e$^-$ $\rightarrow$ Si + OH & 1.500e-07 & -5.000e-01 & 0.000e+00  \\
HSiO$^+$ +     e$^-$ $\rightarrow$ H + SiO & 1.500e-07 & -5.000e-01 &  0.000e+00  \\
C$_5$H$^+$   +    e$^-$  $\rightarrow$ C + C$_4$H & 1.000e-06 & -3.000e-01 & 0.000e+00  \\
C$_5$H$^+$    +   e$^-$ $\rightarrow$ H + C$_5$ & 1.000e-06 & -3.000e-01 & 0.000e+00  \\
HC$_2$N$^+$    +  e$^-$ $\rightarrow$ CH + CN & 1.000e-07 & -5.000e-01 & 0.000e+00  \\
HC$_2$N$^+$     + e$^-$ $\rightarrow$ H + CCN & 1.000e-07 & -5.000e-01 & 0.000e+00  \\
H$_2$C$_4$N$^+$ +    e$^-$ $\rightarrow$ CH + HC$_3$N & 1.000e-07 & -5.000e-01 & 0.000e+00  \\
H$_2$C$_4$N$^+$  +   e$^-$ $\rightarrow$ CCN + C$_2$H$_2$ & 1.000e-07 & -5.000e-01 & 0.000e+00  \\
H$_2$C$_4$N$^+$   +  e$^-$ $\rightarrow$ CN + l-C$_3$H$_2$ & 1.000e-07 & -5.000e-01 &  0.000e+00  \\
HC$_6$N$^+$   +   e$^-$ $\rightarrow$ CN + C$_5$H & 1.000e-07 & -5.000e-01 & 0.000e+00  \\
HC$_6$N$^+$  +    e$^-$ $\rightarrow$ H + C$_6$N & 1.000e-07 & -5.000e-01 & 0.000e+00  \\
HC$_8$N$^+$   +   e$^-$ $\rightarrow$ CN + C$_7$H & 1.000e-07 & -5.000e-01 & 0.000e+00  \\
HC$_8$N$^+$  +    e$^-$ $\rightarrow$ H + C$_8$N & 1.000e-07 & -5.000e-01 & 0.000e+00  \\
HC$_{10}$N$^+$  +   e$^-$ $\rightarrow$ CN + C$_9$H & 1.000e-07 & -5.000e-01 & 0.000e+00  \\
HC$_{10}$N$^+$   +  e$^-$ $\rightarrow$ H + C$_{10}$N & 1.000e-07 & -5.000e-01 & 0.000e+00  \\
GRAIN$^-$    + Cl$^+$ $\rightarrow$ GRAIN0 + Cl & 1.300e-05 & 5.000e-01 & 0.000e+00  \\
GRAIN$^-$  +   F$^+$  $\rightarrow$ GRAIN0 + F & 1.800e-05 & 5.000e-01 & 0.000e+00  \\
\end{longtable}
\clearpage

\section{Differences between the kida.uva.2014 and kida.uva.2024 networks}\label{modif-kida.uva}

The list of species that are new in the kida.uva.2024 network or have been removed as compared to the kida.uva.2014 network are given in Tables~\ref{list-new-species} and ~\ref{list-removed-species}. 

\begin{table*}[h]
    \centering
        \caption{New species in the kida.uva.2024 network.}
    \begin{tabular}{c}
        \hline
        \hline
        Species names \\
        \hline
CNO, HSO, C$_2$N$_2$, HNCS, NCSH, HOCN, HCNO, HONC, HSCN, HCNS, HCCS,\\
CH$_3$S, CH$_2$SH, HC$_3$S, H$_2$CCS, t-C$_3$H$_2$, HC$_3$O, HCCCHO, CH$_3$SH, NH$_2$CHS, \\
H$_2$C$_3$O, c-C$_3$H$_2$O, CH$_2$CCH$_2$, CH$_3$CHS, NH$_2$CH$_2$S, CH$_3$NCO, C$_3$H$_5$, NH$_2$CH$_2$SH,\\ 
C$_2$H$_3$CHO, C$_3$H$_7$, C$_2$H$_5$CHO, C$_3$H$_8$, CS$_2$, H$_2$S$_3$, S$_3$, S$_4$, S$_5$, S$_6$, S$_7$, S$_8$, \\
NHCHO, HOCHN, NH$_2$CO, NH$_2$CH$_2$O, NH$_2$CH$_2$OH, CNCN, NC$_3$N, CH$_3$O, CH$_2$NH$_2$, CH$_3$CHO, \\
C$_4$H$_4$, CH$_3$OCH$_2$, HCOCO, HCOCHO, HOCO, HNCOH$^+$, CSH$^+$, HNCS$^+$, \\
HOCN$^+$, HCNO$^+$, HONC$^+$, HCNS$^+$, HSCN$^+$, H$_2$CSH$^+$, C$_2$N$_2$H$^+$, H$_2$NCS$^+$, HNCSH$^+$, \\
H$_2$NCO$^+$, HNCOH$^+$, H$_2$OCN$^+$, HCNOH$^+$, H$_2$CNO$^+$, H$_2$C$_2$S$^+$, H$_2$SCN$^+$, H$_2$CNS$^+$, \\
HCNSH$^+$, CH$_3$CS$^+$, H$_2$C$_3$S$^+$, C$_2$H$_3$CO$^+$, CH$_3$SH$_2^+$, NH$_2$CHSH$^+$, CH$_3$NCO$^+$, \\
HCCCHOH$^+$, c-C$_3$H$_2$OH$^+$, HC$_5$NH$^+$, CH$_3$CHSH$^+$, CH$_3$NCOH$^+$, NH$_3$CH$_2$SH$^+$, C$_3$H$_7^+$, \\
C$_3$H$_9^+$, H$_5$C$_3$O$^+$, CS$_2^+$, HCS$_2^+$,  CH$_3$SH$^+$ \\
        \hline
    \end{tabular}
    \label{list-new-species}
\end{table*}

\begin{table*}[h]
    \centering
        \caption{Species from kida.uva.2014 removed from the  kida.uva.2024 network.}
    \begin{tabular}{c}
        \hline
        \hline
        Species names.\\
        \hline
H$_2$CSiCH, SiC$_4$H, SiC$_2$CH$_3$, SiC$_6$H, SiC$_3$H$_5$, SiC$_8$H, H$_3$CS$^+$, PNH$_3^+$, H$_3$C$_3$O$^+$, \\
H$_2$C$_5$N$^+$, H$_2$C$_6$N$^+$, H$_2$C$_8$N$^+$, H$_2$C$_{10}$N$^+$\\
        \hline
    \end{tabular}
    \label{list-removed-species}
\end{table*}

\section{References for the rate coefficients}

The references used to update the network are: 
\citet{Adams1976, Adams1984, Adusei1996, Amano1990, Anicich1986, Anicich1993, Anicich2003, Antinolo2016, Antipov2009, Arthur1978, Atkinson2004, Avramenko1967, Azatyan1975, Barckholtz2001, Bauerle1995a, Bauerle1995b, Baulch1994, Baulch2005, Becker2000, Berteloite2010, Bettens1995, Bohland1985, Bohme1988, Biennier2014, Brownsword1997, Bryukov2001, Canosa1995, Carles2011, Carty2006, Chabot2010, Chabot2013, CheikhSidEly2013, Chenel2010, Cohen1991, Cole2012, Derkatch1999, Diau1995, Dombrowsky1992, Ehlerding2004, Eichelberger2007, EpeeEpee2016, Frank1986, Freund1977, Fujii1987, Geppert2004, Gronowski2014, Gustafsson2012, Gustafsson2014, Gustafsson2015, Hamberg2010, Hanson1982, Harada2008, Harada2010, Heays2017, Hemsworth1974, Henshaw1987, Herbst1985, Herbst1987, Herbst1989a, Herbst1989b, Herbst2000, Herbrechtsmeier1973, Hickson2013, Hickson2015, Hickson2016, Hoyermann1996, Humpfer1995, Jensen2000, Kalhori2002, Karkach1999, Kathir2017, Kim1975a, Kim1975b, Leen1988, LeGarrec1997, LeGarrec2003, Lifshitz1991, Lloyd1974, Loison2012, Loison2014a, Loison2014b, Loison2016,  Loison2017, Luca2002, Mackay1981, Mackay1980, Majumdar2017, Maluendes1993, Martinez2008, Martinez2010, Mayer1966, Mayer1967, Mebel1996, Millar1987, Millar1991, Miller1988, Mitchell1984a, Mitchell1984b, Miyoshi1993, Montaigne2005, Morales2010,  Neufeld2005, Neufeld2009, Nguyen2004, Ocana2017, Ostrom2016, OttoR.2008, Paramo2006, Patterson1962, Payzant1975,Peters1973,  Quan2010, Rampino2016, Roberge1991, Rodgers1996, Ruaud2015, Schofield1973, Shin1986, Sims1994, Singleton1988, Smith1981, Smith1984, Smith1992, Smith1994, Snow2009, Stancil1993, Stewart1989, Svensson2015, Szabo2019, Talbi2000, Talbi2009, Thorne1983, Thorne1984, Tizniti2014, Tonolo2020, Tsang1992, Tsang1986, Tsang1991, vanDishoeck1988, Vandooren1994, Vidal2017, Viggiano2005,  Wallington1988, Wakelam2009, Wakelam2015, Warnatz1984, West2019, Woon2009, Yang1993, Yang2011, Zabarnick1989, Zanchet2009, Zellner1988}.

\end{appendix} 

\end{document}